\documentclass[12pt, draftclsnofoot, onecolumn]{IEEEtran}

\usepackage{cite}
\usepackage{amsmath}
\usepackage{amssymb}
\usepackage{graphicx}
\usepackage{epstopdf}
\usepackage{makecell}
\usepackage{color}
\usepackage{subcaption}

\begin{document}
\title{Impartial SWIPT-Assisted User Cooperation Schemes}

\author{Weiyu~Chen,
        Haiyang~Ding, \IEEEmembership{Member,~IEEE,}
        Shilian~Wang, \IEEEmembership{Member,~IEEE,}
        Daniel~Benevides~da~Costa, \IEEEmembership{Senior~Member,~IEEE,}
        and~Fengkui~Gong

\thanks{\scriptsize{W. Chen and S. Wang are with College of Electronic Science and Technology, National University of Defense technology, Changsha, China (email: chenweiyu14@nudt.edu.cn, wangsl@nudt.edu.cn). This work was supported in part by the National Key R\&D Program of China under Grant 2018YFE0100500, in part by the National Natural Science Foundation of China under Grant 61871387, in part by the NUDT Research Fund under Grant ZK17-03-08, and in part by the Natural Science Basic Research Program of Shaanxi under Grant 2019JM-019.}}
\thanks{\scriptsize{H. Ding is with College of Information and Communication, National University of Defense Technology, Xi'an, China (email: dinghy2003@hotmail.com).}}
\thanks{\scriptsize{D. B. da Costa is with the Department of Computer Engineering, Federal University of Cear\'a, Sobral, CE, Brazil (email: danielbcosta@ieee.org).}}
\thanks{\scriptsize{F. Gong is with the State Key Laboratory of Integrated Service Networks, Xidian University, Xi'an, China (e-mail: fkgong@xidian.edu.cn).}}
\thanks{\scriptsize{Corresponding author: Shilian~Wang.}}
}

\maketitle

\IEEEpeerreviewmaketitle

\begin{abstract}
In this paper, we propose an impartial simultaneous wireless information and power transfer (SWIPT)-assisted cooperation mechanism for a non-orthogonal multiple access (NOMA) downlink scenario. Specifically, both a cell-center user and a cell-edge user apply the power-splitting technique and utilize the harvested energy to forward the other user's information on the premise of successful decoding of their own information. Both analytical and numerical results show that the proposed impartial user cooperation mechanism (IUCM) outperforms the traditional partial cooperation mechanism in terms of outage probability, diversity order and diversity-multiplexing trade-off (DMT). For comparison, we further incorporate the IUCM into an orthogonal frequency-division multiple access (OFDMA) framework, which is shown to preserve the same diversity order, while has a worse but more flexible DMT performance in comparison with the IUCM in the NOMA framework. Although the IUCM in OFDMA has a worse outage performance, it is proved that it has the same optimal system outage probability with the IUCM in NOMA when the relaying channel between the two users is error-free.
\end{abstract}

\begin{IEEEkeywords}
Wireless energy harvesting, mutual cooperation, multiple access, outage performance.
\end{IEEEkeywords}

\section{Introduction}
%
%
%
%

\IEEEPARstart{N}{on-orthogonal} multiple access (NOMA) has been recognized as a promising multiple access technique for the fifth-generation (5G) wireless networks to improve spectral efficiency \cite{Dai15CommMag}. Power-domain NOMA, whose key idea is to allocate different power levels to different users using the same time, frequency and code, has drawn significant research interests owing to its low implementation complexity and high compatibility with other techniques \cite{Islam17CommSur}. Specifically, the users within the service area of a base station (BS) are grouped opportunistically. For each group, the users with a weaker channel condition is allocated with a higher power such that the users with a stronger channel condition can subtract the signals of the users with a weaker channel condition from their observations by applying the successive interference cancellation (SIC) technique \cite{Saito13VTC}. In this way, the users remove as much interference as possible and then detect their desired signals. Particularly, the work in \cite{Benjebbour13ISPACS} revealed that NOMA can achieve a 30\% system-level performance improvement than orthogonal multiple access (OMA), while \cite{Sadia18ELEKTRO} investigated the bit error rate of NOMA downlink transmissions under different channel fading types. To further improve the quality-of-service (QoS), the authors in \cite{DingZhiguo15CommuLet} proposed a cooperation scheme in the NOMA framework, which can attain a diversity order of $K$ at all the $K$ users.

Cooperative behaviors will definitely incur additional energy consumption at helping nodes. It is desirable that this energy consumption could be compensated by the source node since it benefits from the cooperative transmission. In this regard, simultaneous wireless information and power transfer (SWIPT), a promising solution to face the tremendous energy consumption growth in the imminent 5G era \cite{huang_ComMagz17,Perera_ComSurTu18,Hu__ComSurTu18}, can be applied between source and relay nodes to solve the problem. Particularly, by simultaneously conducting power transmission and information transmission, SWIPT can improve spectral efficiency, energy efficiency, interference management, and transmission delay \cite{Krikidis_ComMagz14}. The seminal studies of SWIPT implicitly assumed the dual use of the same signals for energy harvesting (EH) and information decoding (ID) \cite{Varshney_ISIT’08,Grover_ISIT’10}, which is not possible until now since a practical EH operation destroys the information carried by the signals. Instead, as summarized in \cite{Krikidis_ComMagz14}, a realistic SWIPT system splits the signals in a certain domain (time, power, antenna or space) to fulfill EH and ID operations simultaneously. To disclose the performance of practical SWIPT systems, for a dual-hop cooperative communication scenario where SWIPT is applied between the source node and the self-sustaining relay node, the authors in \cite{Nasir_TWC’13} proposed two relaying protocols for the time-switching (TS) architecture and power-splitting (PS) architecture, respectively, and characterized their throughputs. Considering a point-to-point transmission scenario, the work in \cite{liu2013TWC} investigated the optimal TS strategies to achieve different trade-offs between EH and ID at the receiver. Furthermore, focusing on a large-scale network where the receivers employ the PS technique, the authors in \cite{Krikidis14TC} studied the trade-off between outage performance and average harvested energy.

Combining NOMA with SWIPT, the work in \cite{Liu16JSAComm} investigated a NOMA downlink scenario where SWIPT is invoked between the BS and cell-center users (relays) to assist the cell-edge users without additional energy consumption at the cell-center users, which dramatically improves the outage performance at the cell-edge users. For the same scenario, an enhanced PS protocol was proposed in \cite{YinghuiYe17ICC}, in which the cell-center users help the cell-edge users on the premise of successful ID of their own information. Based on the gradient decent method, the authors in \cite{Do18SigTelCom} further developed an algorithm to find the optimal PS coefficients at the cell-center users to maximize the sum-throughput. In addition, considering a multiple-input single-output (MISO) NOMA downlink scenario where the cell-center user (relay) adopts a hybrid TS/PS protocol to assist the cell-edge user, the work in \cite{Do17ICC} investigated the performance of different transmit antenna selection criteria at the BS. Different from the studies above, in the SWIPT-assisted NOMA system considered in \cite{Hedayati18TVT}, the harvested energy is used for ID instead of user cooperation, which revealed that NOMA is not always superior to OMA when the energy consumption of ID is taken into account.

Summarizing the foregoing works \cite{Liu16JSAComm,YinghuiYe17ICC,Do18SigTelCom,Do17ICC}, the cell-edge users always act as a weaker, which is helped by the cell-center users opportunistically. Indeed, owing to the dynamic fluctuating characteristics of wireless channels, the instantaneous channel condition between the BS and the cell-edge users can be occasionally better than that between the BS and the cell-center users, which means that the surplus received power at the cell-edge users could also be exploit to help the cell-center users by applying SWIPT between the BS and the cell-edge users, and we refer to this mutual cooperation mechanism between the cell-center users and the cell-edge users as ``\emph{impartial user cooperation mechanism}'' (IUCM), which has the potential to boost the overall transmission performance. Meanwhile, as far as the IUCM is concerned, NOMA's advantage over OMA has not been fully validated yet. To disclose these unknowns, we conduct a comparative study of the proposed IUCM in both the NOMA and the orthogonal frequency-division multiple access (OFDMA) frameworks. The main contributions can be summarized as below:

i) Aiming at taking full advantage of the surplus received power at users and the fluctuating characteristics of wireless channels, an impartial SWIPT-assisted NOMA cooperation (ISANC) protocol is proposed for a classical two-user cooperative NOMA downlink scenario. For such, the outage probability (OP) for both users, system outage probability\footnote{We define the SOP as the probability that either of the two users suffer from information outage.} (SOP), diversity order and diversity-multiplexing trade-off (DMT) of both the ISANC protocol and the conventional SWIPT-assisted NOMA cooperation (CSANC) protocol \cite{YinghuiYe17ICC} are investigated, which shows that the proposed ISANC strictly outperforms CSANC in terms of all the metrics mentioned above.

ii) Invoking the proposed IUCM in an OFDMA framework, an impartial SWIPT-assisted OFDMA cooperation (ISAOC) protocol is presented. Asymptotic analysis shows that the ISAOC protocol has the same diversity order and worse but more flexible DMT performance in comparison with the ISANC protocol. Specifically, for ISAOC, by adjusting the power and frequency allocation, we can continuously enhance the DMT performance at one user at the cost of the DMT performance at the other user. In comparison, for the ISANC protocol, the DMT expressions only have two kinds of forms.

iii) To examine the limiting performance of the proposed protocols, we consider an error-free relaying channel (EFRC), in which case the optimal power allocation for the ISANC protocol and the optimal power and frequency allocation for the ISAOC protocol are derived in terms of SOP. Then, the optimal SOP of the ISANC protocol is proved to be the same as the ISAOC protocol under the EFRC condition.

The rest of the paper is organized as follows: Section II illustrates the system model and describes the three protocols (i.e., CSANC, ISANC and ISAOC). Section III investigates the outage performance, diversity order and DMT performance of the three protocols analytically. The discussion of ISANC and ISAOC under the EFRC scenario is also made in this section. Section IV present the representative numerical results and the corresponding discussion. Finally, Section V concludes the paper.

\section{System Model and the Proposed Impartial Protocols}
\subsection{System Model}
\begin{figure}[!t]
	\centering
	\includegraphics[scale=0.55]{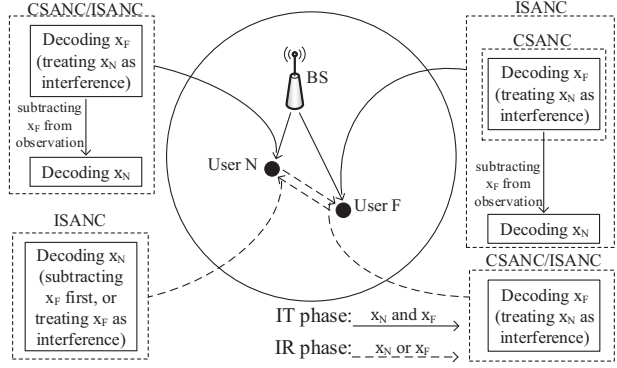}
	\caption{Description of the system model as well as the CSANC and ISANC protocols.}
	\label{Fig_SystemModelIllustration}
\end{figure}

As shown in Fig. \ref{Fig_SystemModelIllustration}, consider a two-user downlink cooperative transmission scenario with two phases, namely, the direct information transmission (IT) phase and the information relaying (IR) phase. During the IT phase, a BS (denoted by B) simultaneously communicates with a cell-center user (denoted by N) and a cell-edge user (denoted by F). The power of the signals transmitted from BS to users N and F are set to $P_{\textrm{N}}$ and $P_{\textrm{F}}$, respectively. Without loss of generality, the spectral efficiency is set to $R$ bit/s/Hz for both users. Utilizing the PS receiver architecture \cite{Krikidis_ComMagz14}, both users can simultaneously conduct EH and ID. Specifically, user N utilizes a fraction of $\beta_{\textrm{N}}$ of its received power from BS for EH and $\left( 1-\beta_{\textrm{N}}\right) $ for ID. Similarly, user F utilizes a fraction of $\beta_{\textrm{F}}$ of its received power from BS for EH and $\left( 1-\beta_{\textrm{F}}\right) $ for ID. All channels are modeled by quasi-static Rayleigh fading (i.e., the channel fading coefficients $h_{\textrm{BN}}$, $h_{\textrm{BF}}$ and $h_{\textrm{NF}}$ pertaining to the B-N, B-F and N-F links remain unchanged during each transmission block but vary independently for different blocks) and perfect channel reciprocity is assumed (i.e., $h_{\textrm{NF}}=h_{\textrm{FN}}$). Accordingly, $|h_{\textrm{BN}}|^2$, $|h_{\textrm{BF}}|^2$ and $|h_{\textrm{NF}}|^2$ obey the exponential distributions with means $\lambda_{\textrm{BN}}$, $\lambda_{\textrm{BF}}$ and $\lambda_{\textrm{NF}}$, respectively. The received signals at N and F from B can be written, respectively, as
\begin{align}
y_{\textrm{BN}}=\left( \sqrt{P_{\textrm{N}}}x_{\textrm{N}}+\sqrt{P_{\textrm{F}}}x_{\textrm{F}} \right)\sqrt{d_{\textrm{BN}}^{-\alpha}}h_{\textrm{BN}}+n_{\textrm{N}},
\label{yBN}
\end{align}
\begin{align}
y_{\textrm{BF}}=\left( \sqrt{P_{\textrm{N}}}x_{\textrm{N}}+\sqrt{P_{\textrm{F}}}x_{\textrm{F}} \right)\sqrt{d_{\textrm{BF}}^{-\alpha}}h_{\textrm{BF}}+n_{\textrm{F}},
\label{yBF}
\end{align}
where $x_{\textrm{N}}$ and $x_{\textrm{F}}$ represent the normalized intended messages for N and F (i.e., $E\left\lbrace |x_{\textrm{N}}|^2 \right\rbrace=E\left\lbrace |x_{\textrm{F}}|^2 \right\rbrace=1 $), $d_{\textrm{BN}}$ and $d_{\textrm{BF}}$ denote the B-N distance and B-F distance, $\alpha$ indicates the path loss exponent, and $n_{\textrm{N}}$ and $n_{\textrm{F}}$ represent the zero-mean additive white Gaussian noise (AWGN) at N and F with variance $\sigma_{\textrm{N}}^2$ and $\sigma_{\textrm{F}}^2$, respectively.

During the IR phase, by consuming the harvested energy in the IT phase, both users can forward the other user's information in a decode-and-forward (DF) mode. Accordingly, the received signal at N from F and that at F from N can be represented, respectively, as 
\begin{align}
y_{\textrm{FN}}=\sqrt{\beta_{\textrm{F}} \eta P_\textrm{B} d_{\textrm{BF}}^{-\alpha} d_{\textrm{NF}}^{-\alpha} |h_{\textrm{BF}}|^2  } 
h_{\textrm{NF}} x_{\textrm{N}} +n_{\textrm{N}},
\label{yFN}
\end{align}
\begin{align}
y_{\textrm{NF}}=\sqrt{\beta_{\textrm{N}} \eta P_\textrm{B} d_{\textrm{BN}}^{-\alpha} d_{\textrm{NF}}^{-\alpha} |h_{\textrm{BN}}|^2  }  
h_{\textrm{NF}} x_{\textrm{F}}+n_{\textrm{F}},
\label{yNF}
\end{align}
where $\eta$ indicates the energy conversion efficiency, $P_\textrm{B} \triangleq P_\textrm{N}+P_\textrm{F}$ denotes the total transmitting power at BS, and $d_{\textrm{NF}}$ represents the N-F distance.

\subsection{The CSANC Protocol}
Fig. \ref{Fig_SystemModelIllustration} also illustrates the main principles of CSANC \cite{YinghuiYe17ICC}. Specifically, during the IT phase, BS broadcasts the superposed signal $\sqrt{P_{\textrm{N}}}x_{\textrm{N}}+\sqrt{P_{\textrm{F}}}x_{\textrm{F}}$ to both user N and user F. User N uses $\beta_{\textrm{N}}$ of its received power for EH and $\left( 1-\beta_{\textrm{N}} \right) $ for ID, while user F utilizes all its received power for ID (i.e., $\beta_{\textrm{F}}\equiv0$). Specifically, according to \eqref{yBN} and \eqref{yBF}, the received signal-to-interference-plus-noise ratio (SINR) at N and F to decode $x_{\textrm{F}}$ can be written, respectively, as\footnote{According to the principles of NOMA, user N decodes $x_{\textrm{F}}$ first, and then subtracts it from its received signal to decode $x_{\textrm{N}}$. In addition, the inequality $P_{\textrm{F}}-P_{\textrm{N}} \left(2^R-1 \right)>0 $ must hold. Otherwise, the minimum required SINR for successfully decoding $x_{\textrm{F}}$ would never be achieved at both users.}
\begin{align}
\gamma_{\textrm{NDF}}^{\textrm{NOMA}}=
\frac{\left(1-\beta_{\textrm{N}} \right)P_\textrm{F} }
{\left(1-\beta_{\textrm{N}} \right)P_\textrm{N} + {d_{\textrm{BN}}^{\alpha} \sigma_{\textrm{N}}^2} / {|h_{\textrm{BN}}|^2}},
\label{SNR_NDF}
\end{align}
\begin{align}
\gamma_{\textrm{FDF}}^{\textrm{NOMA}}=
\frac{\left(1-\beta_{\textrm{F}} \right)P_\textrm{F} }
{\left(1-\beta_{\textrm{F}} \right)P_\textrm{N} + {d_{\textrm{BF}}^{\alpha} \sigma_{\textrm{F}}^2} / {|h_{\textrm{BF}}|^2}}.
\label{SNR_FDF}
\end{align}
Clearly, if $\gamma_{\textrm{FDF}}^{\textrm{NOMA}} \ge 2^R-1$, user F can successfully decode $x_{\textrm{F}}$ during the IT phase, which is denoted by \{FDF=1\} in the rest of this paper. By its turn, \{FDF=0\} is used to indicate that user F can not successfully decode $x_{\textrm{F}}$ during the IT phase. Similarly, \{NDF=1\} denotes that user N can successfully decode $x_{\textrm{F}}$ during the IT phase, in which case the signal-to-noise ratio (SNR) at N to decode $x_{\textrm{N}}$ in the IT phase can be represented as
\begin{align}
\gamma_{\textrm{NDN}}^{\textrm{NOMA}}=
\frac{\left(1-\beta_{\textrm{N}} \right)P_\textrm{N} |h_{\textrm{BN}}|^2} 
{{d_{\textrm{BN}}^{\alpha} \sigma_{\textrm{N}}^2}}.
\label{SNR_NDN}
\end{align}
On the premise of \{NDF=1,NDN=1\}, user N adopts the maximum possible EH factor. Thus, it follows from \eqref{SNR_NDF} and \eqref{SNR_NDN} that its EH factor $\beta_{\textrm{N}}$ can be determined by
\begin{align}
\beta_{\textrm{N}}=\max \left\lbrace 1-\frac{C_{\textrm{N}}^{\textrm{NOMA}}}{|h_{\textrm{BN}}|^2},0\right\rbrace ,
\label{Beta_N}
\end{align}
where
\begin{align}
C_{\textrm{N}}^{\textrm{NOMA}}=
\begin{cases}
\frac{ d_{\textrm{BN}}^{\alpha} \sigma_{\textrm{N}}^2 \left( 2^R-1\right) } {P_{\textrm{N}}},&\textrm{if }k > 2^R,\\
\frac{ d_{\textrm{BN}}^{\alpha} \sigma_{\textrm{N}}^2 \left( 2^R-1\right) } {P_{\textrm{F}}-P_{\textrm{N}} \left( 2^R-1\right) },&\textrm{if }k \le 2^R,
\label{C_N_NOMA}
\end{cases}
\end{align}
in which $k\triangleq P_{\textrm{F}}/P_{\textrm{N}}$ denotes the power allocation ratio.

If \{FDF=0,NDF=1,NDN=1\} happens in the IT phase, during the IR phase, user N will utilize all its harvested energy during the IT phase to forward $x_{\textrm{F}}$ to user F. On the other hand, user F will apply the maximal-ratio combining (MRC) technique to combine the signal from BS (during the IT phase) and the signal from user N (during the IR phase), and attempt to decode $x_{\textrm{F}}$ again. As thus, according to \eqref{yNF} and \eqref{SNR_FDF}, the overall SINR at user F to decode $x_{\textrm{F}}$ in the IR phase can be given by
\begin{align}
\gamma_{\textrm{NHF}}^{\textrm{NOMA}}=
\frac{P_\textrm{F} }
{P_\textrm{N} + {d_{\textrm{BF}}^{\alpha} \sigma_{\textrm{F}}^2} / {|h_{\textrm{BF}}|^2}}
+
\frac{\beta_{\textrm{N}} \eta P_\textrm{B} |h_{\textrm{BN}}|^2 |h_{\textrm{NF}}|^2}
{d_{\textrm{BN}}^{\alpha} d_{\textrm{NF}}^{\alpha} \sigma_{\textrm{F}}^2}.
\label{SNR_NHF}
\end{align}
To proceed, we use \{NHF=1\} to denote that user N successfully helps user F in the IR phase. For the CSANC protocol, user N suffers from information outage (IO) when \{NDF=0\} or \{NDF=1,NDN=0\} happens, while user F suffers from IO when \{FDF=0,NDF=0\}, \{FDF=0,NDF =1,NDN=0\}, or \{NHF=0,FDF=0,NDF=1,NDN=1\} happens.

For CSANC, user N utilizes its received power to assist user F on the premise of successfully decoding user N's own information, which improves the outage performance at F without deteriorating the outage performance at N. Indeed, due to the dynamic fluctuating characteristics of wireless channels, the instantaneous channel condition of the B-N link is not always better than that of the B-F link, which means that user F can also assist user N opportunistically\footnote{Indeed, for NOMA, if the BS has perfect instantaneous channel state information (CSI), it can dynamically allocate power resources such that the user with a worse instantaneous channel condition is always allocated with a higher power level, in which case only the user with a better instantaneous channel condition can assist the other user. However, it is not feasible to provide the BS with precise CSI in most applications due to the high complexity and high overhead, which may cause unaffordable data rate degradation. Moreover, instantaneously acquiring partial CSI still requires a high signaling overhead, especially in a NOMA network with a large number of high mobility users. Alternatively, in this paper, the power allocation at BS is based on the average CSI as in \cite{Liu16JSAComm,YinghuiYe17ICC} (i.e., allocating a higher power lever to the cell-edge users).}. This motivates us to design an impartial cooperation protocol as follows.

\subsection{The ISANC Protocol}
For conciseness, we only describe the differences between CSANC and ISANC, which has been illustrated in Fig. \ref{Fig_SystemModelIllustration}. Specifically, for the ISANC protocol, during the IT phase, user F will also attempt to decode $x_{\textrm{N}}$ with a SIC receiver if the event \{FDF=1\} happens. Accordingly, the SNR at F to decode $x_{\textrm{N}}$ can be represented by
\begin{align}
\gamma_{\textrm{FDN}}^{\textrm{NOMA}}=
\frac{\left(1-\beta_{\textrm{F}} \right)P_\textrm{N} |h_{\textrm{BF}}|^2} 
{{d_{\textrm{BF}}^{\alpha} \sigma_{\textrm{F}}^2}}.
\label{SNR_FDN}
\end{align}
On the premise of \{FDF=1,FDN=1\}, user F adopts the maximum possible EH factor. Therefore, according to \eqref{SNR_FDF} and \eqref{SNR_FDN}, its EH factor $\beta_{\textrm{F}}$ can be determined by
\begin{align}
\beta_{\textrm{F}}=\max \left\lbrace 1-\frac{C_{\textrm{F}}^{\textrm{NOMA}}}{|h_{\textrm{BF}}|^2},0\right\rbrace ,
\label{Beta_F}
\end{align}
where
\begin{align}
C_{\textrm{F}}^{\textrm{NOMA}}=
\begin{cases}
\frac{ d_{\textrm{BF}}^{\alpha} \sigma_{\textrm{F}}^2 \left( 2^R-1\right) } {P_{\textrm{N}}},&\textrm{if }k > 2^R,\\
\frac{ d_{\textrm{BF}}^{\alpha} \sigma_{\textrm{F}}^2 \left( 2^R-1\right) } {P_{\textrm{F}}-P_{\textrm{N}} \left( 2^R-1\right) },&\textrm{if }k \le 2^R. 
\label{C_F_NOMA}
\end{cases}
\end{align}

During the IR phase, if \{NDN=0,FDF=1,FDN=1\} has happened in the IT phase, user F will apply all its harvested energy during the IT phase to forward $x_{\textrm{N}}$ to user N. Accordingly, user N will apply MRC to combine the signal from BS (during the IT phase) and the signal from user F (during the IR phase), and attempt to decode $x_{\textrm{N}}$ again. Note that \{NDN=0,FDF=1,FDN=1\} consists of two mutually exclusive cases. One case is \{NDF=0,FDF=1,FDN=1\}, in which the SINR at user N to decode $x_{\textrm{N}}$ in the IR phase can be given by
\begin{align}
\gamma_{\textrm{FHN}}^{\textrm{NOMA}}=
\frac{P_\textrm{N} }
{P_\textrm{F} + {d_{\textrm{BN}}^{\alpha} \sigma_{\textrm{N}}^2} / {|h_{\textrm{BN}}|^2}}
+
\frac{\beta_{\textrm{F}} \eta P_\textrm{B} |h_{\textrm{BF}}|^2 |h_{\textrm{NF}}|^2}
{d_{\textrm{BF}}^{\alpha} d_{\textrm{NF}}^{\alpha} \sigma_{\textrm{N}}^2}.
\label{SNR_FHN1}
\end{align}
The other case is \{NDF=1,NDN=0,FDF=1,FDN=1\}, in which the SNR at user N to decode $x_{\textrm{N}}$ in the IR phase can be written as
\begin{align}
\gamma_{\textrm{FHN}}^{\textrm{NOMA}}=
\frac{P_\textrm{N} {|h_{\textrm{BN}}|^2}}
{{d_{\textrm{BN}}^{\alpha} \sigma_{\textrm{N}}^2} }
+
\frac{\beta_{\textrm{F}} \eta P_\textrm{B} |h_{\textrm{BF}}|^2 |h_{\textrm{NF}}|^2}
{d_{\textrm{BF}}^{\alpha} d_{\textrm{NF}}^{\alpha} \sigma_{\textrm{N}}^2}.
\label{SNR_FHN2}
\end{align}
Note that according to \eqref{SNR_NDF} and \eqref{SNR_NDN}, \{NDF=1,NDN=0\} happens when $\frac{ d_{\textrm{BN}}^{\alpha} \sigma_{\textrm{N}}^2 \left( 2^R-1\right) } {P_{\textrm{F}}-P_{\textrm{N}} \left( 2^R-1\right) } \le {|h_{\textrm{BN}}|^2} < \frac{ d_{\textrm{BN}}^{\alpha} \sigma_{\textrm{N}}^2 \left( 2^R-1\right) } {P_{\textrm{N}}}$. Therefore, if $ {P_{\textrm{F}}}/{P_{\textrm{N}}}\le 2^R$, the second case \{NDF=1,NDN=0,FDF=1,FDN=1\} will never happen. As before, we use \{FHN=1\} to denote that user F successfully helps user N in the IR phase. For the ISANC protocol, user N suffers from IO in three mutually exclusive events. They are \{NDN=0,FDF=0\}, \{NDN=0,FDF=1,FDN=0\} and \{FHN=0,NDN=0,FDF=1,FDN=1\}.

In the following, to compare the performance of the proposed IUCM in the NOMA framework with that in an OMA framework, the OFDMA-based ISAOC protocol is presented.

\subsection{The ISAOC Protocol}
Similar to the ISANC protocol, for the ISAOC protocol, both users adopt the maximum possible EH factor on the premise of successfully decoding both $x_{\textrm{N}}$ and $x_{\textrm{F}}$ during the IT phase, and then forward the other user's information during the IR phase if possible. The difference lies in that $x_{\textrm{N}}$ and $x_{\textrm{F}}$ are conveyed in orthogonal frequency channels to avoid co-channel interference in the OFDMA framework. As thus, the SNR at user N to decode $x_{\textrm{F}}$ and $x_{\textrm{N}}$ in the IT phase can be written, respectively, as
\begin{align}
\gamma_{\textrm{NDF}}^{\textrm{OFDMA}}=
\frac{\left(1-\beta_{\textrm{N}} \right)P_\textrm{F}{|h_{\textrm{BN}}|^2} }
{ {d_{\textrm{BN}}^{\alpha} \sigma_{\textrm{N}}^2} \theta},
\label{SNR_NDF_O}
\end{align}
\begin{align}
\gamma_{\textrm{NDN}}^{\textrm{OFDMA}}=
\frac{\left(1-\beta_{\textrm{N}} \right)P_\textrm{N}{|h_{\textrm{BN}}|^2} }
{ {d_{\textrm{BN}}^{\alpha} \sigma_{\textrm{N}}^2} \left( 1-\theta\right) },
\label{SNR_NDN_O}
\end{align}
where $\theta \in \left( 0,1\right) $ denotes the percentage of the frequency resources allocated to user F. Note that for ISAOC, \{NDF=1\} happens when $\gamma_{\textrm{NDF}}^{\textrm{OFDMA}} \ge 2^{R/\theta}-1$, while \{NDN=1\} happens when $\gamma_{\textrm{NDN}}^{\textrm{OFDMA}} \ge 2^{R/\left( 1-\theta\right)}-1$. Therefore, the EH factor at user N can be determined by
\begin{align}
\beta_{\textrm{N}}=\max \left\lbrace 1-\frac{C_{\textrm{N}}^{\textrm{OFDMA}}}{|h_{\textrm{BN}}|^2},0\right\rbrace ,
\label{Beta_N_O}
\end{align}
where
\begin{align}
C_{\textrm{N}}^{\textrm{OFDMA}}=
\begin{cases}
&\frac{ d_{\textrm{BN}}^{\alpha} \sigma_{\textrm{N}}^2 \left(1-\theta \right)  } {P_{\textrm{N}}} \left( 2^{\frac{R}{\left(1-\theta \right)}}-1\right),
\textrm{if } \frac{ \left(1-\theta \right)  } {P_{\textrm{N}}} \left( 2^{\frac{R}{\left(1-\theta \right)}}-1\right) > \frac{\theta } {P_{\textrm{F}}} \left( 2^{\frac{R}{\theta} }-1\right),\\
&\frac{ d_{\textrm{BN}}^{\alpha} \sigma_{\textrm{N}}^2 \theta } {P_{\textrm{F}}} \left( 2^{\frac{R}{\theta} }-1\right),
\textrm{if }\frac{ \left(1-\theta \right)  } {P_{\textrm{N}}} \left( 2^{\frac{R}{\left(1-\theta \right)}}-1\right) \le \frac{\theta } {P_{\textrm{F}}} \left( 2^{\frac{R}{\theta} }-1\right). 
\label{C_N_OFDMA}
\end{cases}
\end{align}
Similarly, the SNR at F to decode $x_{\textrm{F}}$ and $x_{\textrm{N}}$ in the IT phase can be written, respectively, as
\begin{align}
\gamma_{\textrm{FDF}}^{\textrm{OFDMA}}=
\frac{\left(1-\beta_{\textrm{F}} \right)P_\textrm{F}{|h_{\textrm{BF}}|^2} }
{ {d_{\textrm{BF}}^{\alpha} \sigma_{\textrm{F}}^2} \theta},
\label{SNR_FDF_O}
\end{align}
\begin{align}
\gamma_{\textrm{FDN}}^{\textrm{OFDMA}}=
\frac{\left(1-\beta_{\textrm{F}} \right)P_\textrm{N}{|h_{\textrm{BF}}|^2} }
{ {d_{\textrm{BF}}^{\alpha} \sigma_{\textrm{F}}^2} \left( 1-\theta\right) }.
\label{SNR_FDN_O}
\end{align}
Accordingly, $\beta_{\textrm{F}}$ can be written as
\begin{align}
\beta_{\textrm{F}}=\max \left\lbrace 1-\frac{C_{\textrm{F}}^{\textrm{OFDMA}}}{|h_{\textrm{BF}}|^2},0\right\rbrace ,
\label{Beta_F_O}
\end{align}
where
\begin{align}
C_{\textrm{F}}^{\textrm{OFDMA}}=
\begin{cases}
&\frac{ d_{\textrm{BF}}^{\alpha} \sigma_{\textrm{F}}^2 \left(1-\theta \right)  } {P_{\textrm{N}}} \left( 2^{\frac{R}{\left(1-\theta \right)}}-1\right),
\textrm{if } \frac{ \left(1-\theta \right)  } {P_{\textrm{N}}} \left( 2^{\frac{R}{\left(1-\theta \right)}}-1\right) > \frac{\theta } {P_{\textrm{F}}} \left( 2^{\frac{R}{\theta} }-1\right),\\
&\frac{ d_{\textrm{BF}}^{\alpha} \sigma_{\textrm{F}}^2 \theta } {P_{\textrm{F}}} \left( 2^{\frac{R}{\theta} }-1\right),
\textrm{if }\frac{ \left(1-\theta \right)  } {P_{\textrm{N}}} \left( 2^{\frac{R}{\left(1-\theta \right)}}-1\right) \le \frac{\theta } {P_{\textrm{F}}} \left( 2^{\frac{R}{\theta} }-1\right). 
\label{C_F_OFDMA}
\end{cases}
\end{align}

During the IR phase, if \{FDF=0,NDF=1,NDN=1\} has happened, user N will forward $x_{\textrm{F}}$ to user F with all its harvested energy in the IT phase. Accordingly, user F will utilize MRC to decode $x_{\textrm{F}}$ again with SNR given by
\begin{align}
\gamma_{\textrm{NHF}}^{\textrm{OFDMA}}=
\frac{P_\textrm{F}{|h_{\textrm{BF}}|^2} }
{{d_{\textrm{BF}}^{\alpha} \sigma_{\textrm{F}}^2} \theta}
+
\frac{\beta_{\textrm{N}} \eta P_\textrm{B} |h_{\textrm{BN}}|^2 |h_{\textrm{NF}}|^2}
{d_{\textrm{BN}}^{\alpha} d_{\textrm{NF}}^{\alpha} \sigma_{\textrm{F}}^2 \theta}.
\label{SNR_NHF_O}
\end{align}
By its turn, if \{NDN=0,FDF=1,FDN=1\} has happened in the IT phase, user F will forward $x_{\textrm{N}}$ to user N in the IR phase and the corresponding SNR at N to decode $x_{\textrm{N}}$ can be written as
\begin{align}
\gamma_{\textrm{FHN}}^{\textrm{OFDMA}}=
\frac{P_\textrm{N}{|h_{\textrm{BN}}|^2} }
{{d_{\textrm{BN}}^{\alpha} \sigma_{\textrm{N}}^2} \left( 1-\theta\right) }
+
\frac{\beta_{\textrm{F}} \eta P_\textrm{B} |h_{\textrm{BF}}|^2 |h_{\textrm{NF}}|^2}
{d_{\textrm{BF}}^{\alpha} d_{\textrm{NF}}^{\alpha} \sigma_{\textrm{N}}^2 \left( 1-\theta\right) }.
\label{SNR_FHN_O}
\end{align}
For the ISAOC protocol, IO happens at N in four mutually exclusive events. They are \{NDN=0, FDF=1,FDN=0\}, \{NDN=0,FDF=0,FDN=1\}, \{NDN=0,FDF=0,FDN=0\} and \{FHN=0,NDN=0, FDF=1,FDN=1\}. On the other hand, IO happens at F when \{FDF=0,NDF=1,NDN=0\}, \{FDF=0, NDF=0,NDN=1\}, \{FDF=0,NDF=0,NDN=0\} or \{NHF=0,FDF=0,NDF=1,NDN=1\} happens.

\section{Performance Analysis and Comparisons}
In this section, we first analyze and compare the performance of the CSANC protocol and the ISANC protocol. Next, we analyze the performance of the ISAOC protocol and compare it with the ISANC protocol. At the end of this section, we study the limiting performance of ISANC and ISAOC by considering an EFRC scenario.
\subsection{Performance Analysis of the CSANC Protocol}
\subsubsection{Analysis of Outage Performance}
According to the analysis of the IO events for the CSANC protocol in Section II, the OP at the cell-center user N can be calculated as
\begin{align}
&P_{\textrm{out,N}}^{\textrm{CSANC}}=
\textrm{Pr}^{\textrm{NOMA}}\left\lbrace \textrm{NDN=0} \right\rbrace
=\textrm{Pr}^{\textrm{NOMA}}\left\lbrace \textrm{NDF=0} \right\rbrace+
\textrm{Pr}^{\textrm{NOMA}}\left\lbrace \textrm{NDF=1,NDN=0} \right\rbrace.
\label{P_N_CSANC}
\end{align}
On the other hand, the OP at the cell-edge user F can be written as
\begin{align}
&P_{\textrm{out,F}}^{\textrm{CSANC}}=
\textrm{Pr}^{\textrm{NOMA}}\left\lbrace \textrm{FDF=0,NDN=0} \right\rbrace
+\textrm{Pr}^{\textrm{NOMA}}\left\lbrace \textrm{NHF=0,FDF=0,NDF=1,NDN=1} \right\rbrace
\nonumber\\
&=\textrm{Pr}^{\textrm{NOMA}}\left\lbrace \textrm{FDF=0} \right\rbrace \times \textrm{Pr}^{\textrm{NOMA}}\left\lbrace \textrm{NDN=0} \right\rbrace 
+\textrm{Pr}^{\textrm{NOMA}}\left\lbrace \textrm{NHF=0,FDF=0,NDF=1,NDN=1} \right\rbrace.
\label{P_F_CSANC}
\end{align}
Furthermore, the SOP can be determined by
\begin{align}
P_{\textrm{sout}}^{\textrm{CSANC}}=&
\textrm{Pr}^{\textrm{NOMA}}\left\lbrace \textrm{NHF=0,FDF=0,NDF=1,NDN=1} \right\rbrace 
+\textrm{Pr}^{\textrm{NOMA}}\left\lbrace \textrm{NDN=0} \right\rbrace.
\label{P_S_CSANC}
\end{align}
Appendix A-1 presents the derivation of all the terms in \eqref{P_N_CSANC}, \eqref{P_F_CSANC} and \eqref{P_S_CSANC}, which lead to an involved expression of $P_{\textrm{sout}}^{\textrm{CSANC}}$ and do not give us any insight. To address this, we further examine the diversity order and the DMT performance of the CSANC protocol as follows.

\subsubsection{Diversity Order and DMT}\quad

\textbf{\emph{Proposition 1}}: The achievable diversity order of the CSANC protocol is 1.

\emph{Proof}: According to \eqref{Pr_NDF=0_NOMA_HighSNR}, \eqref{Pr_NDF=1,NDN=0_NOMA_HighSNR}, \eqref{Pr_NHF=0_NOMA_HighSNR3}, \eqref{P_N_CSANC} and \eqref{P_S_CSANC}, as $P_{\textrm{B}} \to \infty$, $P_{\textrm{sout}}^{\textrm{CSANC}}$ decays proportional to $1/P_{\textrm{B}}$, which completes the proof.\hfill$\blacksquare$
 
On the other hand, the DMT, a trade-off between diversity order and normalized spectral efficiency (NSE, a.k.a., multiplexing gain), is widely used to evaluate the comprehensive performance of communication systems \cite{Zheng03TIT,Laneman04TIT,Ding12TWC}. Specifically, NSE is defined as the ratio of the actual spectral efficiency to the maximum achievable spectral efficiency, while diversity order is known as the decaying rate of the OP in high SNR region and is a function of NSE. The NSE at the cell-center user N can be written as
\begin{align}
r_{\textrm{N}}={R}/{\log \left(1+\frac{P_{\textrm{B}} \lambda_{\textrm{BN}}}{d_{\textrm{BN}}^{\alpha} \sigma_{\textrm{N}}^2} \right) }.
\label{MultiplexingGain_N}
\end{align}
Similarly, user F's NSE can be represented as
\begin{align}
r_{\textrm{F}}={R}/{\log \left(1+\frac{P_{\textrm{B}} \lambda_{\textrm{BF}}}{d_{\textrm{BF}}^{\alpha} \sigma_{\textrm{F}}^2} \right) }.
\label{MultiplexingGain_F}
\end{align}
As mentioned in Section II, the inequality $k>2^R-1$ must hold due to the principle of NOMA. Nonetheless, it can be observed from \eqref{MultiplexingGain_N} and \eqref{MultiplexingGain_F} that with a fixed NSE in the DMT analysis, $R$ approaches to infinity as $P_{\textrm{B}}$ approaches to infinity. Thus, $k$ has to vary in the DMT analysis to satisfy the inequality. To address this, by incorporating a scaling factor $a$ and a displacement factor $b$, we formulate the term $k$ in a general form as
\begin{align}
k=a\left( 2^R-1\right) +b,
\label{k}
\end{align}
where $a=1,b>0$ or $a>1,b\ge0$.

\textbf{\emph{Proposition 2}}: For the CSANC protocol, the DMT of user N is $\left( 1-2r_{\textrm{N}} \right) $. If $a>1$, the DMT of user F is $\left( 2-3r_{\textrm{F}}\right) $. Otherwise, user F's DMT is $\left( 2-4r_{\textrm{F}}\right) $.

\emph{Proof}: Please refer to Appendix A-3.\hfill$\blacksquare$

\emph{Remark 1}: It follows from Proposition 2 that the gap between the power allocation ratio $k$ and the threshold $(2^R-1)$ plays a critical role in the DMT performance. Interestingly, when the scaling factor $a$ is greater than 1, the achievable multiplexing gain (AMG) for user F will be enhanced from 1/2 to 2/3 for CSANC. This means that in the practical system setup, we should maintain a higher scaling factor $a$ to boost the multiplexing gain of the cell-edge user.

\emph{Corollary 1}: For the CSANC protocol, the achievable diversity order of user N is 1, while user F's achievable diversity order is 2.

\emph{Proof}: The proof is straightforward as per Proposition 2.\hfill$\blacksquare$ 

\emph{Remark 2}: The higher achievable diversity order at user F than the counterpart at user N can be explained by the fact that the CSANC protocol provides a redundant information transfer path for the cell-edge user F (i.e., the N-F link). Nonetheless, as can be observed from Proposition 1 and Corollary 1, since user N can only reach a diversity order of one, the overall achievable diversity order is constrained to one. This results from the fact that there is no redundant information transfer path for the cell-center user N except for the B-N link for the CSANC protocol.

\subsection{Performance Analysis of the ISANC Protocol}
\subsubsection{Analysis of Outage Performance}
For the ISANC protocol, the OP with respect to user F is the same as that in the CSANC protocol, which can be written as
\begin{align}
P_{\textrm{out,F}}^{\textrm{ISANC}}=P_{\textrm{out,F}}^{\textrm{CSANC}}=&
\textrm{Pr}^{\textrm{NOMA}}\left\lbrace \textrm{FDF=0} \right\rbrace \times \textrm{Pr}^{\textrm{NOMA}}\left\lbrace \textrm{NDN=0} \right\rbrace \nonumber\\
&+\textrm{Pr}^{\textrm{NOMA}}\left\lbrace \textrm{NHF=0,FDF=0,NDF=1,NDN=1} \right\rbrace.
\label{P_F_ISANC}
\end{align}
Different from the CSANC protocol, the cell-center user N is likely to be assisted by the cell-edge user F in ISANC. According to the analysis of the IO events for the ISANC protocol in Section II, user N's OP can be expressed as
\begin{align}
P_{\textrm{out,N}}^{\textrm{ISANC}}=&
\textrm{Pr}^{\textrm{NOMA}}\left\lbrace \textrm{FDN=0} \right\rbrace \times \textrm{Pr}^{\textrm{NOMA}}\left\lbrace \textrm{NDN=0} \right\rbrace 
+\textrm{Pr}^{\textrm{NOMA}}\left\lbrace \textrm{FHN=0,NDF=0,FDF=1,FDN=1} \right\rbrace\nonumber\\
&+\textrm{Pr}^{\textrm{NOMA}}\left\lbrace \textrm{FHN=0,NDF=1,NDN=0,FDF=1,FDN=1} \right\rbrace,
\label{P_N_ISANC}
\end{align}
where
\begin{align}
&\textrm{Pr}^{\textrm{NOMA}}\left\lbrace \textrm{FDN=0} \right\rbrace 
=\textrm{Pr}^{\textrm{NOMA}}\left\lbrace \textrm{FDF=0} \right\rbrace+
\textrm{Pr}^{\textrm{NOMA}}\left\lbrace \textrm{FDF=1,FDN=0} \right\rbrace.
\label{P_FDN_NOMA}
\end{align}
Next, for the ISANC protocol, SOP can be determined by
\begin{align}
P_{\textrm{sout}}^{\textrm{ISANC}}=&
\textrm{Pr}^{\textrm{NOMA}}\left\lbrace \textrm{FDN=0} \right\rbrace \times \textrm{Pr}^{\textrm{NOMA}}\left\lbrace \textrm{NDN=0} \right\rbrace 
+\textrm{Pr}^{\textrm{NOMA}}\left\lbrace \textrm{FHN=0,NDF=0,FDF=1,FDN=1} \right\rbrace\nonumber\\
&+\textrm{Pr}^{\textrm{NOMA}}\left\lbrace \textrm{FHN=0,NDF=1,NDN=0,FDF=1,FDN=1} \right\rbrace\nonumber\\
&+\textrm{Pr}^{\textrm{NOMA}}\left\lbrace \textrm{NHF=0,FDF=0,NDF=1,NDN=1} \right\rbrace.
\label{P_S_ISANC}
\end{align}

\textbf{\emph{Proposition 3}}: For the ISANC protocol, the OP at the cell-center user N and the SOP are smaller than or at most the same as the counterparts for the CSANC protocol.

\emph{Proof}: \eqref{P_N_ISANC} can be rewritten as
\begin{align}
P_{\textrm{out,N}}^{\textrm{ISANC}}=&
\textrm{Pr}^{\textrm{NOMA}}\left\lbrace \textrm{NDN=0} \right\rbrace \times \left\lbrace  
\textrm{Pr}^{\textrm{NOMA}}\left\lbrace \textrm{FDN=0} \right\rbrace \right.   
\left. + \textrm{Pr}^{\textrm{NOMA}}\left\lbrace   \textrm{FHN=0,FDF=1,FDN=1}|\textrm{NDN=0} \right\rbrace  \right\rbrace \nonumber\\
&\le \textrm{Pr}^{\textrm{NOMA}}\left\lbrace \textrm{NDN=0} \right\rbrace=P_{\textrm{out,N}}^{\textrm{CSANC}}.
\end{align}
On the other hand, \eqref{P_S_ISANC} can be rewritten as
\begin{align}
P_{\textrm{sout}}^{\textrm{ISANC}}=&
\textrm{Pr}^{\textrm{NOMA}}\left\lbrace \textrm{NHF=0,FDF=0,NDF=1,NDN=1} \right\rbrace 
+
\textrm{Pr}^{\textrm{NOMA}}\left\lbrace \textrm{NDN=0} \right\rbrace \nonumber\\
&\times \left\lbrace  
\textrm{Pr}^{\textrm{NOMA}}\left\lbrace \textrm{FDN=0} \right\rbrace \right.   \left. + \textrm{Pr}^{\textrm{NOMA}}\left\lbrace   \textrm{FHN=0,FDF=1,FDN=1}|\textrm{NDN=0} \right\rbrace  \right\rbrace \nonumber\\
&\le 
\textrm{Pr}^{\textrm{NOMA}}\left\lbrace \textrm{NHF=0,FDF=0,NDF=1,NDN=1} \right\rbrace 
+\textrm{Pr}^{\textrm{NOMA}}\left\lbrace \textrm{NDN=0} \right\rbrace=P_{\textrm{sout}}^{\textrm{CSANC}},
\end{align}
which completes the proof.\hfill$\blacksquare$ 

\emph{Remark 3}: From equation \eqref{P_F_ISANC} and Proposition 3, we can observe that the ISANC protocol preserves the outage performance at the cell-edge user F in the CSANC protocol, while improves both the outage performance at N and the overall outage performance (i.e., SOP). This benefits from the fact that the proposed IUCM constructs one more information transfer path (i.e., the F-N link) for the cell-center user N such that the surplus received power at both user N and user F and the fluctuating characteristics of both the B-N link and B-F link are fully exploited.

Appendix A-1 presents the derivation of all the terms in \eqref{P_F_ISANC}, \eqref{P_N_ISANC} and \eqref{P_S_ISANC}, based on which we examine the diversity order and the DMT performance of the ISANC protocol as follows.

\subsubsection{Diversity Order and DMT}\quad

\textbf{\emph{Proposition 4}}: The achievable diversity order of the ISANC protocol is 2.

\emph{Proof}: According to Appendix A-2, as $P_{\textrm{B}} \to \infty$, $P_{\textrm{sout}}^{\textrm{ISANC}}$ decays proportional to $1/P_{\textrm{B}}^2$ since all terms in \eqref{P_S_ISANC} decay proportional to $1/P_{\textrm{B}}^2$, which completes the proof.\hfill$\blacksquare$

\textbf{\emph{Proposition 5}}: For the ISANC protocol, the DMT of user N is $\left( 2-4r_{\textrm{N}} \right) $. If $a>1$, the DMT of user F is $\left( 2-3r_{\textrm{F}}\right) $. Otherwise, user F's DMT is $\left( 2-4r_{\textrm{F}}\right) $.

\emph{Proof}: Please refer to Appendix A-3.\hfill$\blacksquare$ 

\emph{Remark 4}: It can be observed from Propositions 2 and 5 that the proposed ISANC protocol outperforms the CSANC protocol in terms of DMT performance at N, while the two protocols have the same DMT performance at F. This again shows that the ISANC protocol can improve the QoS at N without deteriorating the QoS at F, which results from the fact that user F only uses its surplus received power to additionally forward $x_{\textrm{N}}$ on the premise of successfully decoding its own information in the proposed ISANC protocol. In addition, it is worth noticing that similar to the CSANC protocol, the AMG for user F will be increased from 1/2 to 2/3 for the ISANC protocol when the scaling factor $a$ is greater than 1.

\emph{Corollary 2}: For the ISANC protocol, both users N and F can achieve a diversity order of 2.

\emph{Proof}: The proof is straightforward as per Proposition 5.\hfill$\blacksquare$ 

\emph{Remark 5}: It can be observed from Propositions 1 and 4 as well as Corollaries 1 and 2 that the achievable diversity order at the cell-center user N is enhanced from one to two in the ISANC protocol, leading thus to a better overall diversity performance than CSANC.

\subsection{Performance Analysis of the ISAOC Protocol}
\subsubsection{Analysis of Outage Performance}
According to the analysis of the IO events for the ISAOC protocol in Section II, the OP with respect to the cell-edge user F can be represented as
\begin{align}
P_{\textrm{out,F}}^{\textrm{ISAOC}}=&
 \textrm{Pr}^{\textrm{OFDMA}}\left\lbrace \textrm{NHF=0,FDF=0,NDF=1,NDN=1} \right\rbrace \nonumber\\
&+\textrm{Pr}^{\textrm{OFDMA}}\left\lbrace \textrm{FDF=0} \right\rbrace \times \left(1- \textrm{Pr}^{\textrm{OFDMA}}\left\lbrace \textrm{NDF=1,NDN=1} \right\rbrace\right).
\label{P_F_ISAOC}
\end{align}
Different from ISANC, user N and user F are peers in ISAOC since there is no requirement of decoding order. Therefore, the analysis of the two users is absolutely symmetrical, and the OP at the cell-center user N can be given by
\begin{align}
P_{\textrm{out,N}}^{\textrm{ISAOC}}=&
\textrm{Pr}^{\textrm{OFDMA}}\left\lbrace \textrm{FHN=0,NDN=0,FDF=1,FDN=1} \right\rbrace \nonumber\\
&+\textrm{Pr}^{\textrm{OFDMA}}\left\lbrace \textrm{NDN=0} \right\rbrace \times \left(1- \textrm{Pr}^{\textrm{OFDMA}}\left\lbrace \textrm{FDF=1,FDN=1} \right\rbrace\right).
\label{P_N_ISAOC}
\end{align}
Furthermore, the SOP can be determined by
\begin{align}
P_{\textrm{sout}}^{\textrm{ISAOC}}=&
\textrm{Pr}^{\textrm{OFDMA}}\left\lbrace \textrm{NHF=0,FDF=0,NDF=1,NDN=1} \right\rbrace \nonumber\\
&+\textrm{Pr}^{\textrm{OFDMA}}\left\lbrace \textrm{FHN=0,NDN=0,FDF=1,FDN=1} \right\rbrace\nonumber\\
&+
\textrm{Pr}^{\textrm{OFDMA}}\left\lbrace \textrm{FDF=0} \right\rbrace \times \left(1- \textrm{Pr}^{\textrm{OFDMA}}\left\lbrace \textrm{NDF=1,NDN=1} \right\rbrace\right)\nonumber\\
&+\textrm{Pr}^{\textrm{OFDMA}}\left\lbrace \textrm{NDN=0} \right\rbrace \times \textrm{Pr}^{\textrm{OFDMA}}\left\lbrace \textrm{FDF=1,FDN=0} \right\rbrace.
\label{P_S_ISAOC}
\end{align}
Appendix B-1 presents the derivation of all the terms in \eqref{P_F_ISAOC}, \eqref{P_N_ISAOC} and \eqref{P_S_ISAOC}, based on which we examine the diversity order and the DMT performance of the ISAOC protocol as follows.
\subsubsection{Diversity Order and DMT}\quad

\textbf{\emph{Proposition 6}}: The achievable diversity order of the ISAOC protocol is 2.

\emph{Proof}: According to Appendix B-2, as $P_{\textrm{B}} \to \infty$, $P_{\textrm{sout}}^{\textrm{ISAOC}}$ decays proportional to $1/P_{\textrm{B}}^2$ since all terms in \eqref{P_S_ISAOC} decay proportional to $1/P_{\textrm{B}}^2$, which completes the proof.\hfill$\blacksquare$

\textbf{\emph{Proposition 7}}: For the ISAOC protocol, if 
$\frac{ \left(1-\theta \right)  } {P_{\textrm{N}}} \left( 2^{\frac{R}{\left(1-\theta \right)}}-1\right) > \frac{\theta } {P_{\textrm{F}}} \left( 2^{\frac{R}{\theta} }-1\right)$
, the DMTs at user N and user F are 
$\left( 2-\frac{2r_{\textrm{N}}}{1-\theta} \right)$
 and $\min\left\lbrace2-\frac{r_{\textrm{F}}}{\theta\left( 1-\theta\right) }, 2-\frac{2r_{\textrm{F}}}{\theta }\right\rbrace$
 , respectively. Otherwise, the DMTs at N and F are $\min\left\lbrace2-\frac{r_{\textrm{N}}}{\theta\left( 1-\theta\right) }, 2-\frac{2r_{\textrm{N}}}{\left( 1-\theta \right) }\right\rbrace$ and $\left( 2-\frac{2r_{\textrm{F}}}{\theta} \right)$, respectively.

\emph{Proof}: Please refer to Appendix B-3.\hfill$\blacksquare$ 

\emph{Remark 6}: It can be observed from Proposition 7 that for the ISAOC protocol, the AMGs at user N and user F are $\left( 1-\theta\right)$ and $\min\left\lbrace 2\theta\left( 1-\theta\right),\theta \right\rbrace$, respectively, when $\frac{ \left(1-\theta \right) } {P_{\textrm{N}}} \left( 2^{\frac{R}{\left(1-\theta \right)}}-1\right) > \frac{\theta } {P_{\textrm{F}}} \left( 2^{\frac{R}{\theta} }-1\right)$. In this case, at the cost of the AMG at F, we can enhance the AMG at N by allocating more frequency resources to user N. Similarly, when $\frac{ \left(1-\theta \right)  } {P_{\textrm{N}}} \left( 2^{\frac{R}{\left(1-\theta \right)}}-1\right) \le \frac{\theta } {P_{\textrm{F}}} \left( 2^{\frac{R}{\theta} }-1\right)$, the AMGs at user N and user F are $\min\left\lbrace 2\theta\left( 1-\theta\right),\left( 1-\theta \right) \right\rbrace$ and $\theta$, respectively. In this case, by allocating more frequency resources to user F, we can increase the AMG at F at the expense of the AMG at N. Overall, to maximize the minimum AMG at N and F, $\theta$ should equal to 0.5, leading thus to the AMGs of 1/2 at both N and F. Comparing these results with Proposition 5, we can conclude that ISANC outperforms ISAOC in terms of DMT, since we can achieve an AMG of 1/2 at N and an AMG of 2/3 at F simultaneously in ISANC. However, as can be observed from Propositions 7 and 5, by adjusting the power and frequency allocation, the ISAOC protocol can adjust the AMGs at both users continuously, which benefits from the flexibility of resources allocation in OFDMA. In comparison, the DMT performance of the ISANC protocol only has two kinds of forms, which is due to the the fact that NOMA does not involve frequency resources allocation and has a power allocation constraint (i.e., $k>2^R-1$).

\emph{Corollary 3}: For the ISAOC protocol, both users N and F can achieve a diversity order of 2.

\emph{Proof}: The proof is straightforward as per Proposition 7.\hfill$\blacksquare$ 

\emph{Remark 7}: It can be observed from Propositions 1, 4 and 6 as well as Corollaries 1, 2 and 3 that in comparison with the conventional partial cooperation mechanism in CSANC, IUCM would indeed improve the achievable diversity order at the cell-center user N, leading thus to a higher achievable diversity order of two for both ISANC and ISAOC.
\subsection{Discussion Under EFRC Scenario}
In this part, to examine the limiting performance of the proposed ISANC and ISAOC protocols, we discuss their performance under the EFRC scenario, which can be accomplished by connecting user N and user F with an ideal optical fiber or assuming that the inter-user cooperation are accomplished perfectly with the aid of genius. Specifically, under the EFRC scenario, we first analyze the optimal power allocation for ISANC to obtain its optimal SOP. Next, we derive the optimal power and frequency allocation for ISAOC and then determine its optimal SOP. We will find out that the optimal SOPs of the two protocols are the same.
\subsubsection{Optimal SOP of the ISANC Protocol}
Under the EFRC scenario, user F can definitely recover $x_{\textrm{F}}$ when user N forward $x_{\textrm{F}}$ to user F, and vice versa. In such an ideal case, the SOP of the ISANC protocol can be given by 
\begin{align}
&P_{\textrm{isout}}^{\textrm{ISANC}}=
\textrm{Pr}^{\textrm{NOMA}}\left\lbrace \textrm{NDN=0} \right\rbrace +\textrm{Pr}^{\textrm{NOMA}}\left\lbrace \textrm{FDN=0} \right\rbrace.
\label{P_ideaS_ISANC}
\end{align}

\textbf{\emph{Proposition 8}}: Under the EFRC scenario, the optimal power allocation for the ISANC protocol is $k=P_{\textrm{F}}/P_{\textrm{N}}=2^R$ and the corresponding optimal SOP can be given by
 \begin{align}
 P_{\textrm{isout,opt}}^{\textrm{ISANC}}=&\left(1-\exp \left(-\frac{d_{\textrm{BN}}^{\alpha} \sigma_{\textrm{N}}^{2}}{\lambda_{\textrm{BN}} P_{\textrm{B}}}\left(2^{2 R}-1\right)\right)\right)
 \times
 \left(1-\exp \left(-\frac{d_{\textrm{BF}}^{\alpha} \sigma_{\textrm{F}}^{2}}{\lambda_{\textrm{BF}} P_{\textrm{B}}}\left(2^{2 R}-1\right)\right)\right).
 \label{P_ideaS_opt_ISANC}
 \end{align}

\emph{Proof}: 
By combining \eqref{P_ideaS_ISANC} and \eqref{Pr_NDF=0_NOMA}$\sim$\eqref{Pr_FDF=1,FDN=0_NOMA}, we can arrive at the analytical expression of $P_{\textrm{isout}}^{\textrm{ISANC}}$, which turns out to be a monotonically increasing function of $k$ when $k>2^R$ and a monotonically decreasing function of $k$ when $k\le2^R$. As a result, the minimum $P_{\textrm{isout}}^{\textrm{ISANC}}$ is achieved when $k=2^R$. Next, by replacing $P_{\textrm{F}}$ and $P_{\textrm{N}}$ with $\frac{2^{\textrm{R}}}{1+2^{\textrm{R}}} P_{\textrm{B}}$ and $\frac{1}{1+2^{\textrm{R}}} P_{\textrm{B}}$, respectively, in the derived expression of $P_{\textrm{isout}}^{\textrm{ISANC}}$, we can arrive at \eqref{P_ideaS_opt_ISANC}, which completes the proof.
\hfill$\blacksquare$ 
\subsubsection{Optimal SOP of the ISAOC Protocol}
Under the EFRC scenario, both users can definitely recover the information sent from each other such that the SOP of the ISAOC protocol can be written as
\begin{align}
P_{\textrm{isout}}^{\textrm{ISAOC}}=&\left(1-\textrm{Pr}^{\textrm{OFDMA}}\left\lbrace \textrm{FDF=1,FDN=1}\right\rbrace\right)
\times
\left(1-\textrm{Pr}^{\textrm{OFDMA}}\left\lbrace \textrm{NDF=1,NDN=1}\right\rbrace\right).
\label{P_ideaS_ISAOC}
\end{align}

\textbf{\emph{Proposition 9}}: Under the EFRC scenario, the optimal power allocation for the ISAOC protocol can be expressed as $\frac{(1-\theta)}{P_{\textrm{N}}}\left(2^{\frac{R}{1-\theta}}-1\right)=\frac{\theta}{P_{\textrm{F}}}\left(2^{\frac{R}{\theta}}-1\right)$, based on which the optimal frequency allocation is $\theta=0.5$, and the corresponding optimal SOP can be given by \eqref{P_ideaS_opt_ISANC}.

\emph{Proof}: 
It follows from \eqref{P_ideaS_ISAOC}, \eqref{Pr_NDF=1,NDN=1_OFDMA}, \eqref{Pr_FDF=1,FDN=1_OFDMA}, \eqref{C_N_OFDMA} and \eqref{C_F_OFDMA} that $P_{\textrm{isout}}^{\textrm{ISAOC}}$ is a monotonically increasing function of $k$ when $\frac{(1-\theta)}{P_{\textrm{N}}}\left(2^{\frac{R}{1-\theta}}-1\right)>\frac{\theta}{P_{\textrm{F}}}\left(2^{\frac{R}{\theta}}-1\right)$, whereas it is a monotonically decreasing function of $k$ when $\frac{(1-\theta)}{P_{\textrm{N}}}\left(2^{\frac{R}{1-\theta}}-1\right)\le \frac{\theta}{P_{\textrm{F}}}\left(2^{\frac{R}{\theta}}-1\right)$. Therefore, the minimum $P_{\textrm{isout}}^{\textrm{ISAOC}}$ is achieved on the premise of power allocation satisfying to $\frac{(1-\theta)}{P_{\textrm{N}}}\left(2^{\frac{R}{1-\theta}}-1\right)=\frac{\theta}{P_{\textrm{F}}}\left(2^{\frac{R}{\theta}}-1\right)$, in which case it can be rewritten as 
$P_{\textrm{isout}}^{\textrm{ISAOC}}=\left(1-e^{-\frac{d_{\textrm{BF}}^{\alpha} \sigma_{\textrm{F}}^{2}}{\lambda_{\textrm{BF}} P_{\textrm{B}}} f\left(\theta \right)}\right)\times
\left(1-e^{-\frac{d_{\textrm{BN}}^{\alpha} \sigma_{\textrm{N}}^{2}}{\lambda_{\textrm{BN}} P_{\textrm{B}}}f\left(\theta \right)}\right)$, where $f\left(\theta \right) \triangleq \theta\left(2^{\frac{R}{\theta}}-1\right)+(1-\theta)\left(2^{\frac{R}{1-\theta}}-1\right)$. Next, the proof can be completed by observing that $f'\left(0.5 \right)=0$ and $f''\left(\theta \right)>0$ ($\theta\in(0,1)$).
\hfill$\blacksquare$ 

\emph{Corollary 4}: Under the EFRC scenario, the ISANC protocol and the ISAOC protocol have the same optimal SOP.

\emph{Proof}: The proof is straightforward according to Propositions 8 and 9.\hfill$\blacksquare$ 

\emph{Remark 8}: The conclusion in Corollary 4 can be explained by the fact that when the relaying channel is error-free, the two users can be regarded as one user, and the overall power and frequency resources occupied by this ``user'' in ISANC is essentially the same as that in ISAOC.
\section{Numerical Results and Discussion}
In this section, representative numerical results are provided to validate the theoretical analysis in Section III and to compare the performance of the three protocols. Without loss of generality, the system parameters setup is adopted as $R=1$bit/s/Hz, $P_{\textrm{B}}=20$dBm, $\sigma_{\textrm{F}}^2=\sigma_{\textrm{N}}^2=-50$dBm, $\eta=0.5$, $\alpha=2$, $\lambda_{\textrm{BF}}=\lambda_{\textrm{BN}}=\lambda_{\textrm{NF}}=1$, $d_{\textrm{BF}}=35$m, $d_{\textrm{BN}}=25$m and $d_{\textrm{NF}}=10$m unless otherwise specified. The power allocation ratio ($k$) is set to 7/3 for NOMA, while the percentage of power resources as well as the percentage of the frequency resources that is allocated to user F are set to 50\% for OFDMA unless otherwise specified.

\begin{figure*}[!t]
	\centering
	\begin{minipage}[t]{0.32\textwidth}
		\includegraphics[scale=0.41]{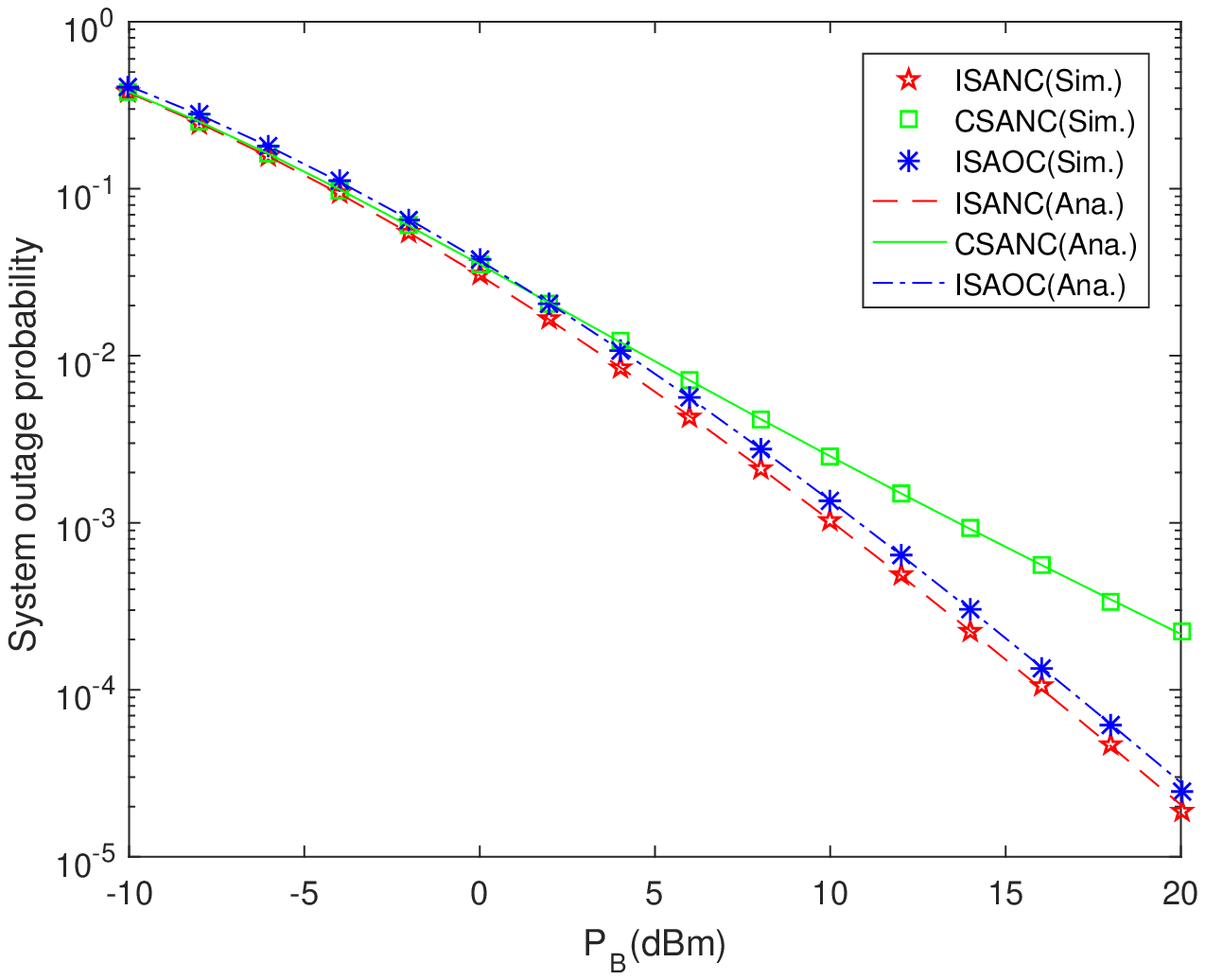} 
		\subcaption{}
	\end{minipage}
	\hfil
	\begin{minipage}[t]{0.32\textwidth}
		\includegraphics[scale=0.41]{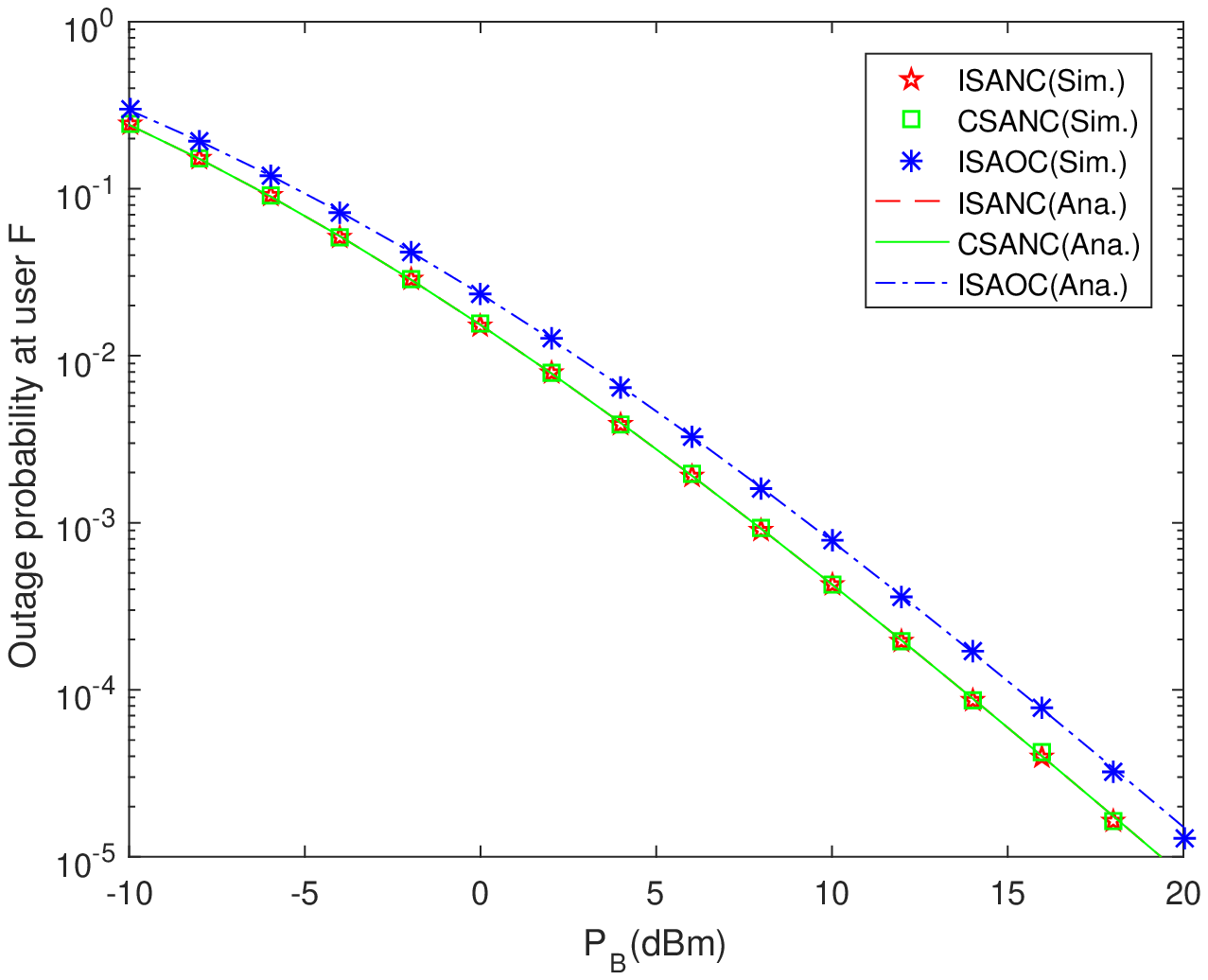} 
		\subcaption{}
	\end{minipage}
	\hfil
	\begin{minipage}[t]{0.32\textwidth}
		\includegraphics[scale=0.41]{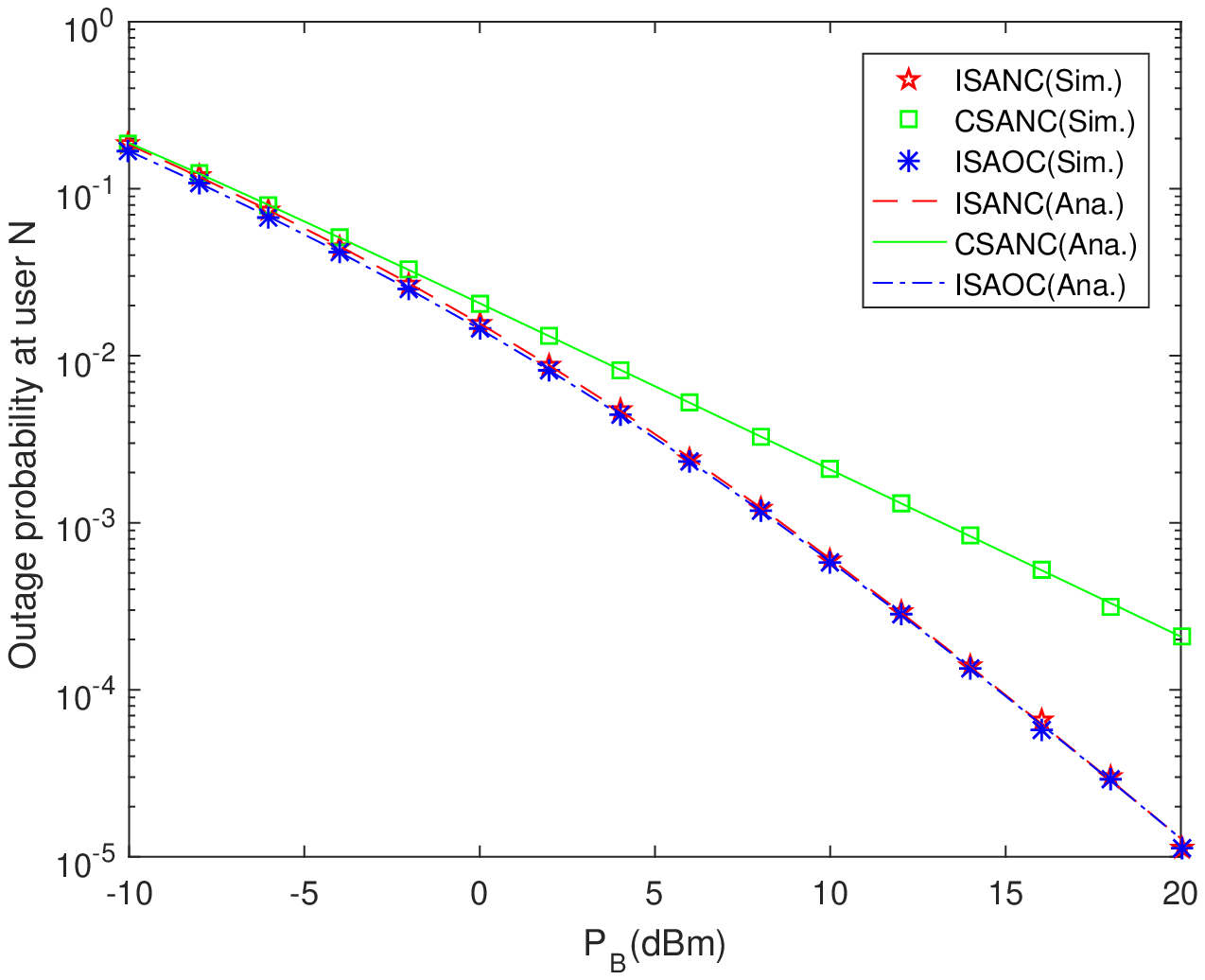}
		\subcaption{}
	\end{minipage}
	\caption{Outage performance of the three protocols. (a) SOP, (b) OP at the cell-edge user F, (c) OP at the cell-center user N.}
	\label{Fig_Validation}
\end{figure*}
Fig. \ref{Fig_Validation} validates the theoretical results developed in Section III. Specifically, as can be observed, the simulation results of the outage performance of the three protocols match well with the derived analytical expressions in Appendixes A-1 and B-1. Particularly, the ISANC protocol and the CSANC protocol have the same OP at user F, while the OP at user N and the SOP of the ISANC protocol are better than the counterparts of the CSANC protocol, which complies with Proposition 3. In addition, the ISANC protocol and the ISAOC protocol have the same decaying rate of SOP, which is faster than the CSANC protocol. This complies with Propositions 1, 4 and 6. On the other hand, the three protocols have the same decaying rate of the OP at the cell-edge user F. In comparison, only the ISANC protocol and ISAOC protocol have the same decaying rate of the OP at the cell-center user N, which is again better than the CSANC protocol. These observations complies with Corollaries 1, 2 and 3.

\begin{figure*}[!t]
	\centering
	\begin{minipage}[t]{0.32\textwidth}
		\includegraphics[scale=0.41]{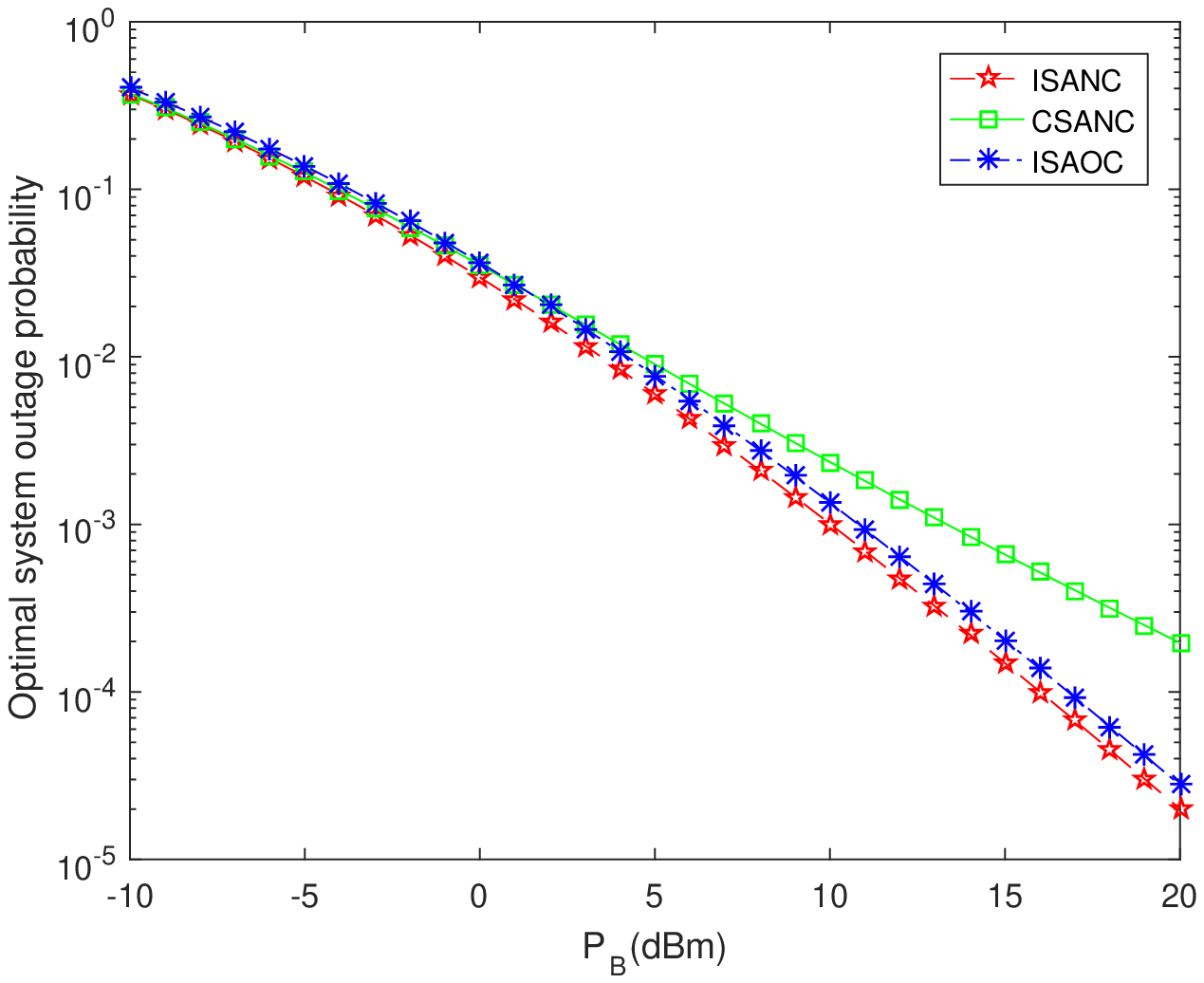} 
		\subcaption{}
	\end{minipage}
	\hfil
	\begin{minipage}[t]{0.32\textwidth}
		\includegraphics[scale=0.41]{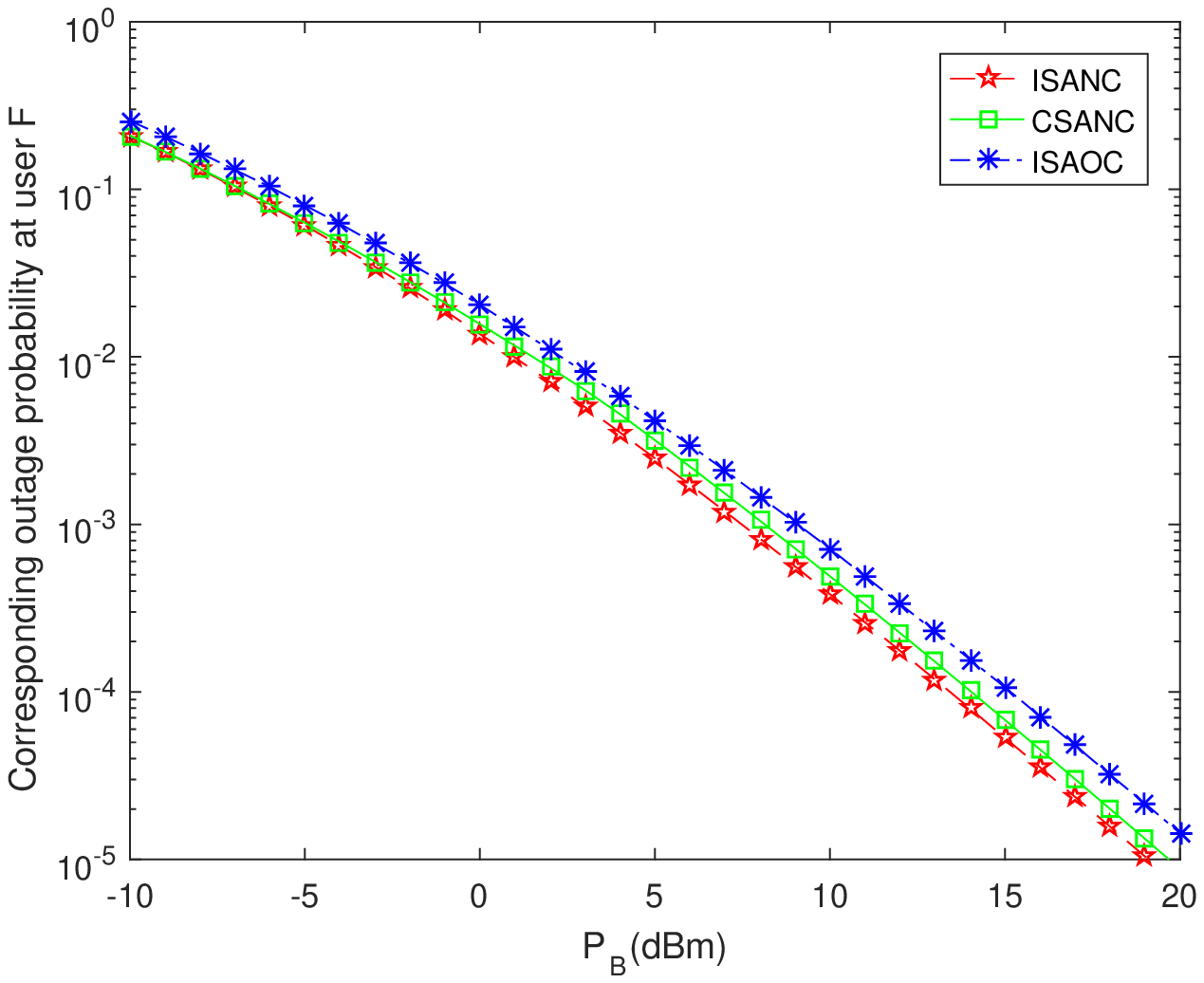} 
		\subcaption{}
	\end{minipage}
	\hfil
	\begin{minipage}[t]{0.32\textwidth}
		\includegraphics[scale=0.41]{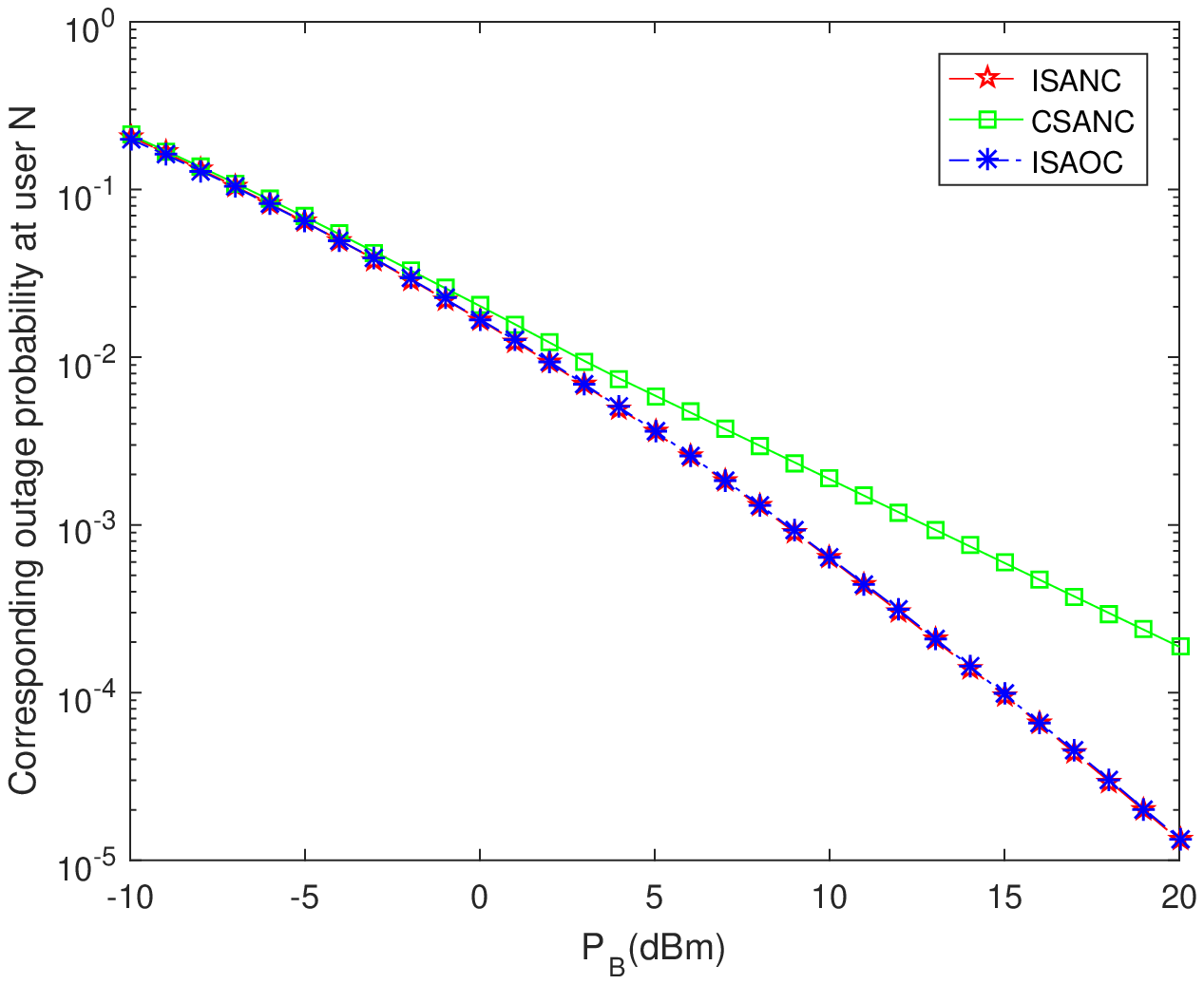}
		\subcaption{}
	\end{minipage}
	\caption{Comparison of the optimal SOP of the three protocols. (a) The optimal SOP, (b) Corresponding OP at the cell-edge user F, (c) Corresponding OP at the cell-center user N.}
	\label{Fig_OptVersusPB}
\end{figure*}
Fig. \ref{Fig_OptVersusPB} presents the optimal SOP and the corresponding OPs at users N and F for the three protocols, in which the optimal SOP is acquired via violent traverse of power and frequency allocation (herein only the ISAOC protocol involves frequency allocation). Several observations are drawn as follows: 1) The ISANC protocol outperforms the ISAOC protocol in terms of the optimal SOP, which agrees with our knowledge about NOMA's superiority over OMA \cite{Benjebbour13ISPACS}; 2) The ISANC protocol outperforms the CSANC protocol in terms of the optimal SOP, which is straightforward since Proposition 3 has shown that the SOP of the ISANC protocol is always smaller than or at most the same as that of the CSANC protocol; 3) The SOP of the CSANC protocol is mainly limited by the OP at user N. In comparison, both the ISANC and the ISAOC protocols have more balanced OPs at N and F when adopting the optimal power and frequency allocation (herein ``optimal allocation'' is calculated in terms of the SOP), which again manifests the advantage of the IUCM over the conventional partial cooperation mechanism.

\begin{figure*}[!t]
	\centering
	\begin{minipage}[t]{0.32\textwidth}
		\includegraphics[scale=0.41]{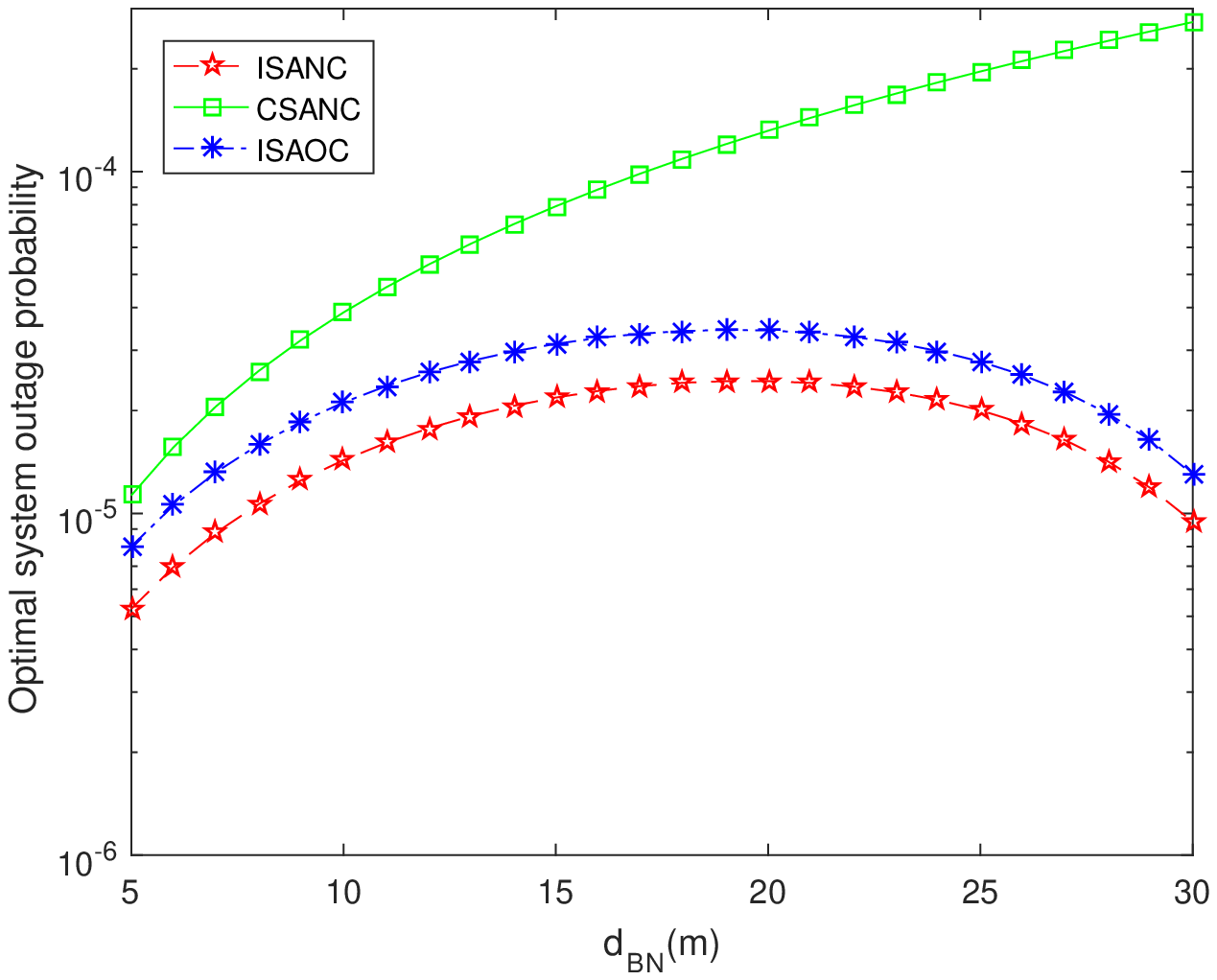} 
		\subcaption{}
	\end{minipage}
	\hfil
	\begin{minipage}[t]{0.32\textwidth}
		\includegraphics[scale=0.41]{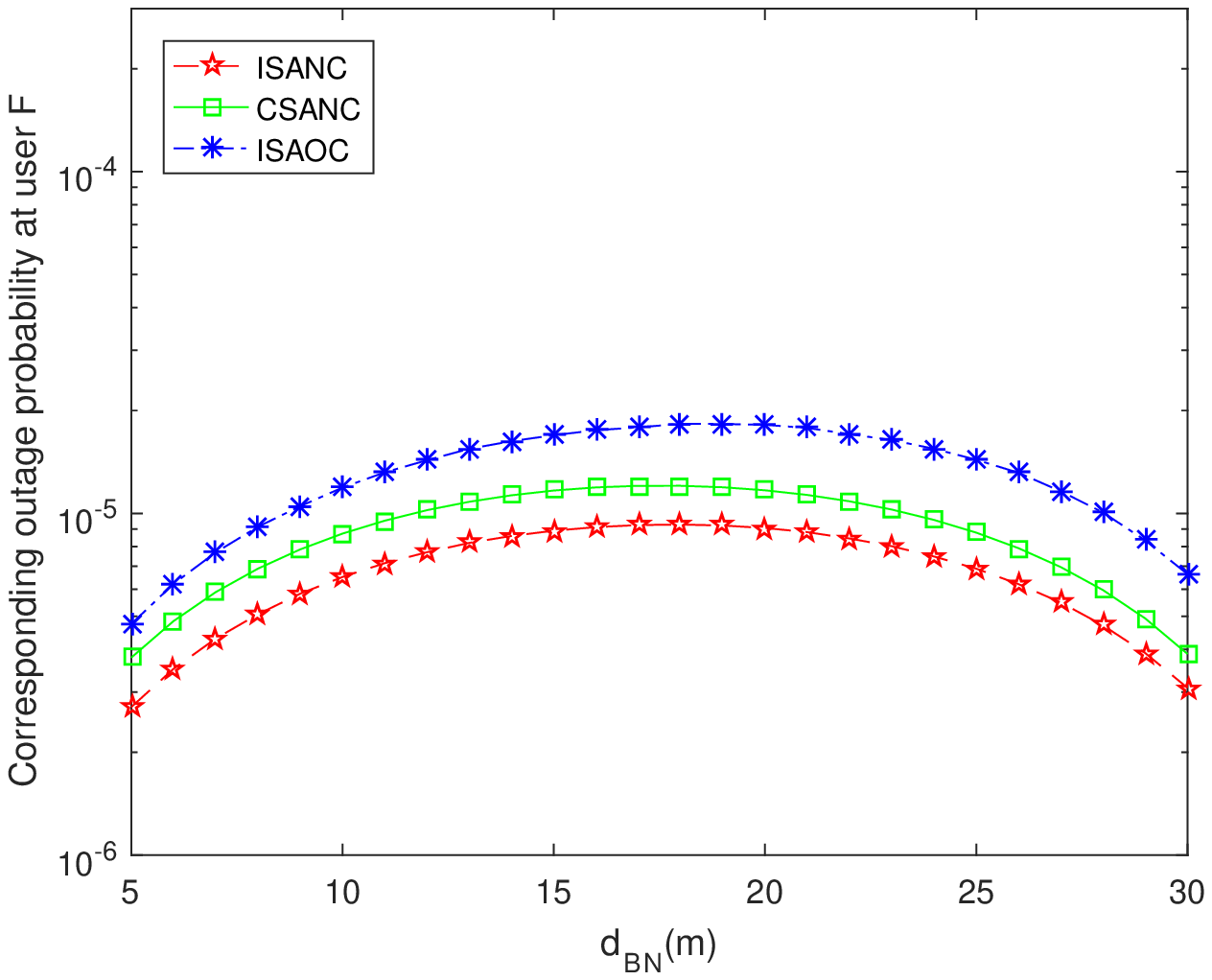} 
		\subcaption{}
	\end{minipage}
	\hfil
	\begin{minipage}[t]{0.32\textwidth}
		\includegraphics[scale=0.41]{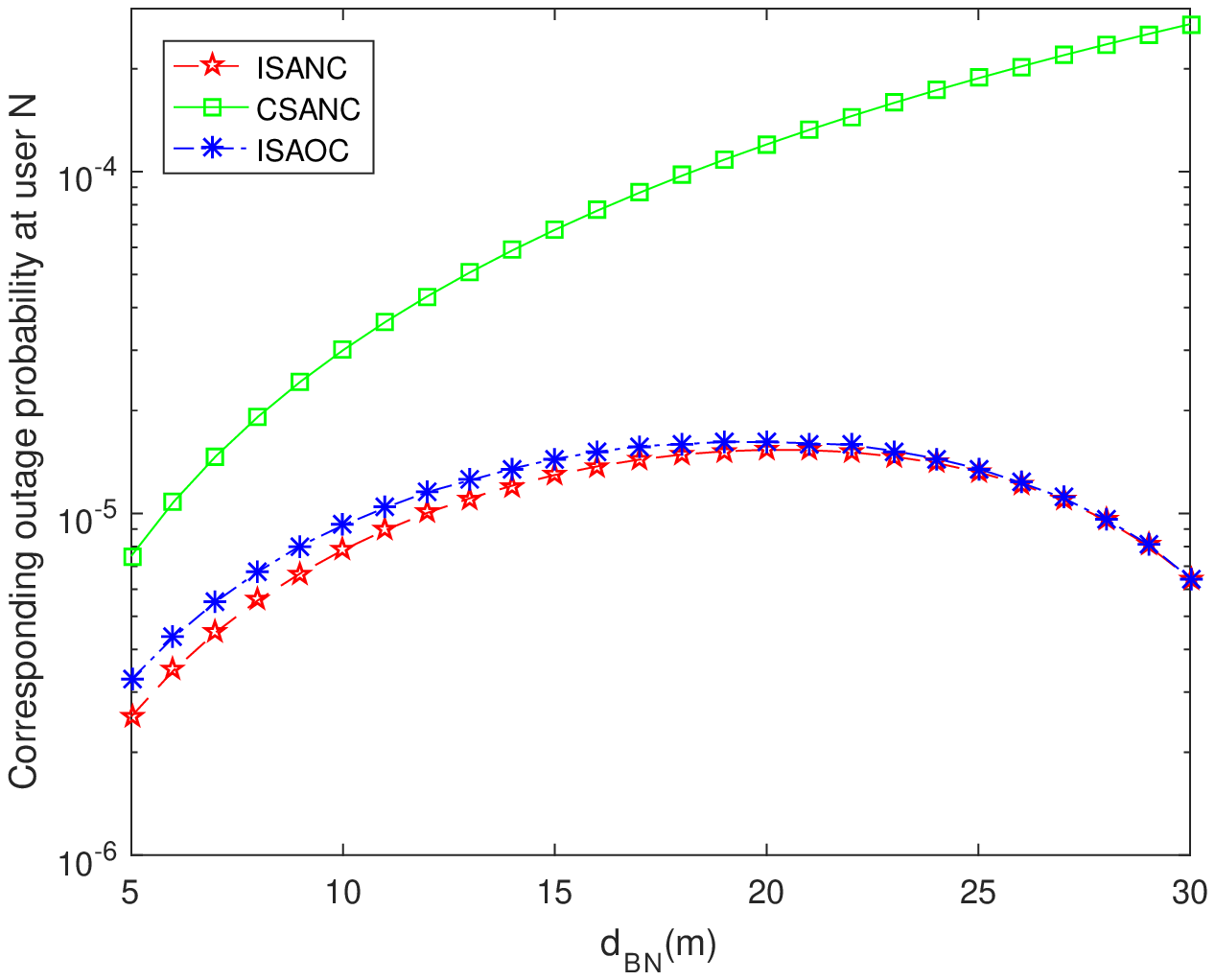}
		\subcaption{}
	\end{minipage}
	\caption{Effects of the location of the cell-center user N on the optimal SOP of the three protocols. (a) The optimal SOP, (b) Corresponding OP at the cell-edge user F, (c) Corresponding OP at the cell-center user N.}
	\label{Fig_OptVersus_d}
\end{figure*}
Fig. \ref{Fig_OptVersus_d} shows the effects of the location of the cell-center user N on the optimal SOP and the corresponding OPs at users N and F for the three protocols. Specifically, in the figure, we set $d_{\textrm{BF}}=35$m and $d_{\textrm{NF}}=d_{\textrm{BF}}-d_{\textrm{BN}}$, and the optimal SOP is calculated via violent traverse of power and frequency allocation as before. Note that Fig. \ref{Fig_OptVersus_d} validates all the three observations drawn from Fig. \ref{Fig_OptVersusPB}. Herein, additional observations are drawn from Figure 4 as below: 1) For both the ISANC protocol and the ISAOC protocol, the SOP first increase and then decrease with the B-N distance. This can be explained by the fact that when user N is close to BS, it can readily recover the information from BS so that the BS can allocate more resources to user F, leading thus to a good outage performance at both users. On the other hand, when user N is close to user F, their mutual cooperation becomes more effective, which results in an improvement of the outage performance at both users; 2) In comparison, for the CSANC protocol, the OP at N increases monotonically with the B-N distance, leading thus to a continuous increase in the SOP. This is due to the fact that the cell-center user N can not get assistance from the cell-edge user F in the traditional partial cooperation mechanism; 3) When adopting the optimal power and frequency allocation, the ISANC protocol outperforms the ISAOC protocol in terms of the OPs at both user N and user F. However, as user N gets closer to user F, the OP gap at N for the two protocols decreases; 4) For all the three protocols, the SOP is higher when user N is close to BS instead of user F. This observation may give us some guidance about user pairing.

\begin{figure}
	\centering
		\centering
	\begin{minipage}[t]{0.46\textwidth}
		\includegraphics[scale=0.5]{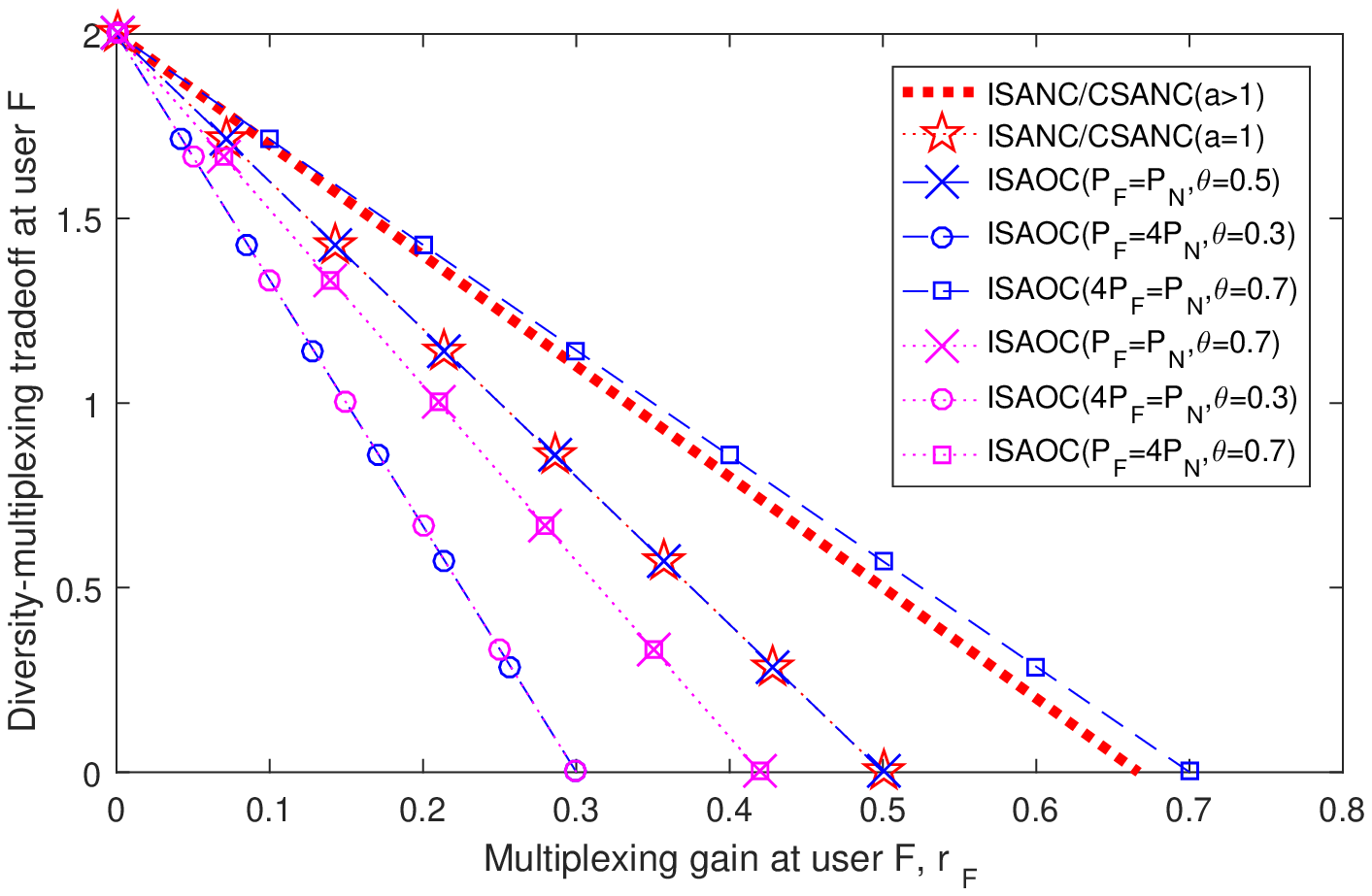} 
		\subcaption{}
	\end{minipage}
	\begin{minipage}[t]{0.46\textwidth}
		\includegraphics[scale=0.5]{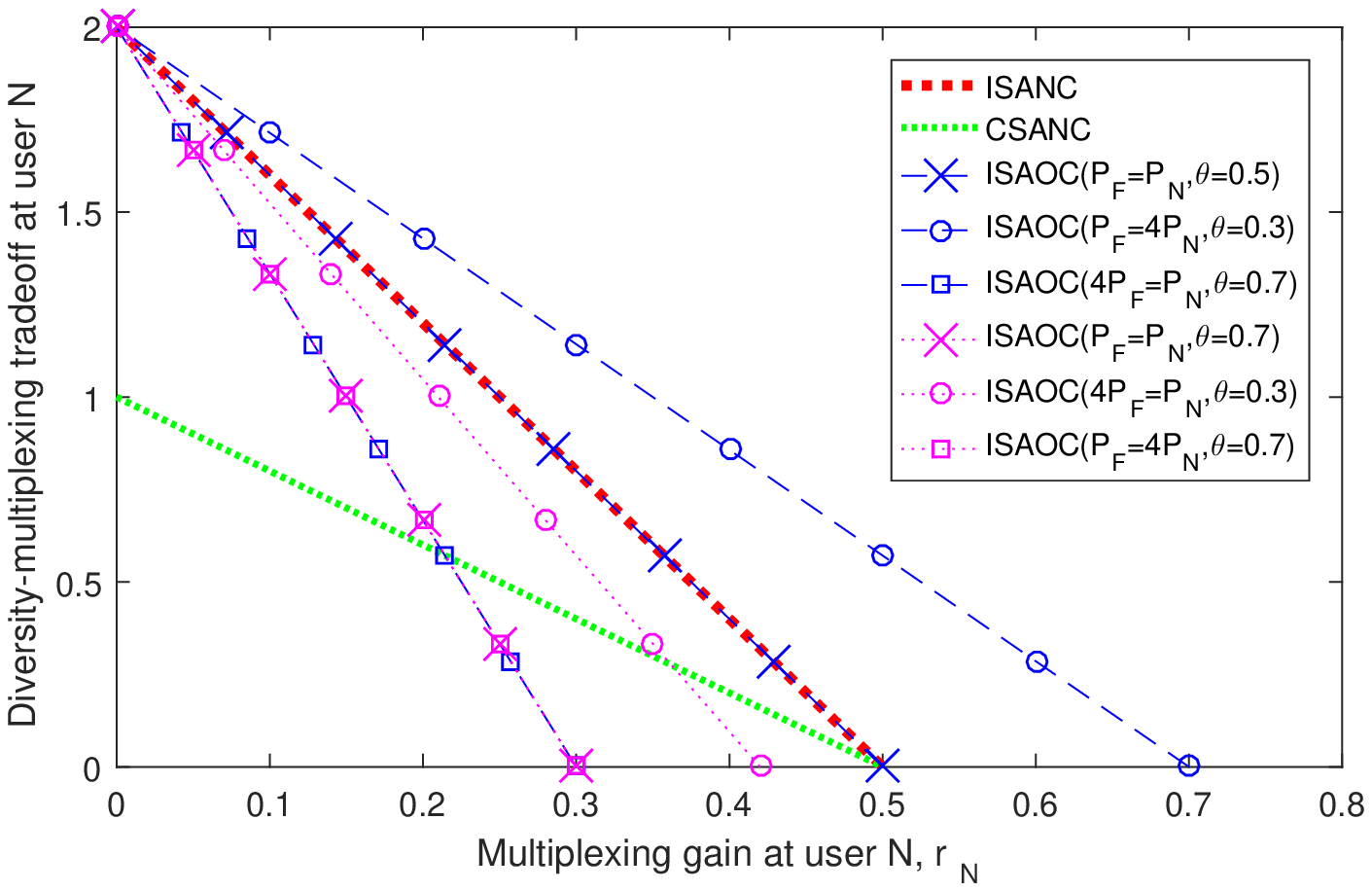} 
		\subcaption{}
	\end{minipage}
\caption{Comparison of the three protocols' DMT performance at: (a) The cell-edge user F, (b) The cell-center user N.}
\label{Fig_DMT}
\end{figure}
Fig. \ref{Fig_DMT} illustrates the DMT performance of the three protocols, in which the red dotted lines represent the optimal DMT performance of the ISANC protocol. As stated in Section III, for the ISAOC protocol, we can flexibly adjust the DMT performance at the two users by changing the power and frequency allocation. Specifically, as can be observed from the figure, the max-min AMG of 1/2 can be achieved at both the cell-center user N and the cell-edge user F by allocating equal frequency resources to them. Furthermore, we can increase the AMG at N by 1/5 at the cost of a 1/5 decrease in the AMG at F by allocating more frequency resources but less power resources to N, and vice versa. Note that such a high AMG (7/10) at one user is unachievable for ISANC, which benefits from the flexibility of resources allocation\footnote{As stated in Section III, the ISANC protocol does not involve frequency allocation and its power allocation is constrained by $k>2^R-1$.} in ISAOC. In addition, as can be observed from the magenta dotted lines in the figure, compared with equal frequency allocation, improper resources allocation in ISAOC may lead to a decrease in the AMGs at both user N and user F.

\begin{figure}[!t]
	\centering
	\includegraphics[scale=0.47]{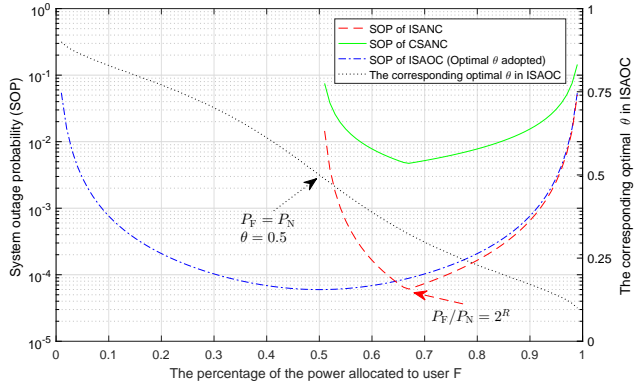}
	\caption{Comparison of the three protocols' SOP under the EFRC scenario.}
	\label{Fig_EFRC}
\end{figure}
Fig. \ref{Fig_EFRC} validates our analytical results under the EFRC scenario, in which the optimal SOP of the ISAOC protocol and the corresponding optimal $\theta$ are presented for each power allocation setup via violence search of frequency allocation. Several observations are drawn as follows: 1) For ISANC, the optimal power allocation (i.e., $k=2^R$) complies with Proposition 8; 2) For ISAOC, the optimal joint power and frequency allocation (i.e., $P_\textrm{F}=P_\textrm{N},\theta=0.5$) matches with Proposition 9; 3) The optimal SOPs of ISANC and ISAOC are the same, which agrees with Corollary 4; 4) The optimal SOP of ISANC is much better than that of CSANC. This is due to the fact that for the IUCM, the cell-center user N can get assistance from the cell-edge user F, which is an effective approach to improve the SOP, especially under the EFRC scenario.

\section{Concluding Remarks}
In this paper, the ISANC protocol was proposed to fully utilize the dynamic fluctuating characteristics of wireless channels in SWIPT-assisted NOMA systems, which enables not only the cell-center user but also the cell-edge user to utilize the surplus received power to assist the other user on the premise of successfully decoding their own information. Both analytical and numerical results manifested that the proposed IUCM outperforms the conventional partial cooperation mechanism in terms of OP, diversity order and DMT.

Furthermore, to compare the performance of the IUCM in the NOMA framework with that in the OFDMA framework, the OFDMA-based ISAOC protocol was presented. Compared with ISANC, ISAOC has the same diversity order but worse DMT. Nonetheless, for ISAOC, by adjusting the power and frequency allocation, we can continuously improve the DMT at one user at the cost of that at the other user, while the DMT expressions of ISANC only have two kinds of forms. Interestingly, although ISANC has a better outage performance, when the relaying channel is error-free, it was proved that the two protocols have the same optimal SOP.

\appendices
\section{}
\setcounter{equation}{0}
\renewcommand{\theequation}{\thesection.\arabic{equation}}
\textbf{A-1: Probability of the Events in the NOMA Framework}

For conciseness, we denote $|h_{\textrm{BF}}|^{2}$, $|h_{\textrm{BN}}|^{2}$ and $|h_{\textrm{NF}}|^{2}$ by $x$, $y$ and $z$, respectively, in the rest of the paper. According to \eqref{SNR_NDF}, we can arrive at
\begin{align}
&\textrm{Pr}^{\textrm{NOMA}}\left\lbrace \textrm{NDF=0} \right\rbrace=
\int_{0}^{\frac{ d_{\textrm{BN}}^{\alpha} \sigma_{\textrm{N}}^2 \left( 2^R-1\right) } {P_{\textrm{F}}-P_{\textrm{N}} \left( 2^R-1\right) }}
\frac{1}{\lambda_{\textrm{BN}}}e^{\frac{-y}{\lambda_{\textrm{BN}}}} dy
=1-e^{-\frac{ d_{\textrm{BN}}^{\alpha} \sigma_{\textrm{N}}^2 \left( 2^R-1\right) } {\lambda_{\textrm{BN}} \left( P_{\textrm{F}}-P_{\textrm{N}} \left( 2^R-1\right) \right) }}.
\label{Pr_NDF=0_NOMA}
\end{align}
In the same way, it follows from \eqref{SNR_FDF} that
\begin{align}
\textrm{Pr}^{\textrm{NOMA}}\left\lbrace \textrm{FDF=0}\right\rbrace =1-e^{-\frac{d_{\textrm{BF}}^{\alpha} \sigma_{\textrm{F}}^{2}\left(2^{R}-1\right)}{\lambda_{\textrm{BF}}\left(P_{\textrm{F}}-P_{\textrm{N}}\left(2^{\textrm{R}}-1\right)\right)}}.
\label{Pr_FDF=0_NOMA}
\end{align}
Next, it follows from \eqref{SNR_NDF} and \eqref{SNR_NDN} that when $k \le 2^R$, $\textrm{Pr}^{\textrm{NOMA}}\left\lbrace \textrm{NDF=1,NDN=0}\right\rbrace=0$ holds. Otherwise, we have
\begin{align}
&\textrm{Pr}^{\textrm{NOMA}}\left\lbrace \textrm{NDF=1,NDN=0}\right\rbrace  
=
e^{-\frac{d_{\textrm{BN}}^{\alpha} \sigma_{\textrm{N}}^{2}\left(2^{R}-1\right)}{\lambda_{\textrm{BN}}\left(P_{\textrm{F}}-P_{\textrm{N}}\left(2^{\textrm{R}}-1\right)\right)}} 
-e^{-\frac{d_{\textrm{BN}}^{\alpha} \sigma_{\textrm{N}}^{2}\left(2^{R}-1\right)}{\lambda_{\textrm{BN}} P_{\textrm{N}}}}.
\label{Pr_NDF=1,NDN=0_NOMA}
\end{align}
Similarly, according to \eqref{SNR_FDF} and \eqref{SNR_FDN}, when $k \le 2^R$, $\textrm{Pr}^{\textrm{NOMA}}\left\lbrace \textrm{FDF=1,FDN=0}\right\rbrace=0$ holds. Otherwise, one can show that
\begin{align}
&\textrm{Pr}^{\textrm{NOMA}}\left\lbrace \textrm{FDF=1,FDN=0}\right\rbrace 
=
e^{-\frac{d_{\textrm{BF}}^{\alpha} \sigma_{\textrm{F}}^{2}\left(2^{R}-1\right)}{\lambda_{\textrm{BF}}\left(P_{\textrm{F}}-P_{\textrm{N}}\left(2^{\textrm{R}}-1\right)\right)}} 
-e^{-\frac{d_{\textrm{BF}}^{\alpha} \sigma_{\textrm{F}}^{2}\left(2^{R}-1\right)}{\lambda_{\textrm{BF}} P_{\textrm{N}}}}.
\label{Pr_FDF=1,FDN=0_NOMA}
\end{align}
Furthermore, it follows from \eqref{SNR_NDF}$\sim$\eqref{Beta_N} and \eqref{SNR_NHF} that
\begin{align}
&\textrm{Pr}^{\textrm{NOMA}}\left\lbrace \textrm{NHF=0,FDF=0,NDF=1,NDN=1} \right\rbrace\nonumber\\
&=\textrm{Pr}\left\lbrace x<\frac{ d_{\textrm{BF}}^{\alpha} \sigma_{\textrm{F}}^2 \left( 2^R-1\right) } {P_{\textrm{F}}-P_{\textrm{N}} \left( 2^R-1\right) } ,
y\ge C_{\textrm{N}}^{\textrm{NOMA}},\right.   
\left.   \frac{P_{\textrm{F}} x}{P_{\textrm{N}} x+ d_{\textrm{BF}}^{\alpha} \sigma_{\textrm{F}}^2}+\frac{\eta P_{\textrm{B}} z \left(y-C_{\textrm{N}}^{\textrm{NOMA}} \right) }{d_{\textrm{BN}}^{\alpha} d_{\textrm{NF}}^{\alpha} \sigma_{\textrm{F}}^2}<\left( 2^R-1\right)\right\rbrace\nonumber\\
&=\int_{0}^{\frac{d_{\textrm{BF}}^{\alpha} \sigma_{\textrm{F}}^{2}\left(2^{R}-1\right)}{P_{\textrm{F}}-P_{\textrm{N}}\left(2^{R}-1\right)}}
\int_{C_{\textrm{N}}^{\textrm{NOMA}}}^{\infty}
\int_{0}^{\frac{\left(2^{R}-1\right)-\frac{P_{\textrm{F}}}{P_{\textrm{N}}+\frac{d_{\textrm{BF}}^{\alpha} \sigma_{\textrm{F}}^{2}}{x}}}{\frac{\eta P_{\textrm{B}}\left(y-C_{\textrm{N}}^{\textrm{NOMA}}\right)}{d_{\textrm{BN}}^{\alpha} d_{\textrm{NF}}^{\alpha} \sigma_{\textrm{F}}^{2}}}}
\frac{1}{\lambda_{\textrm{BF}}}e^{-\frac{x}{\lambda_{\textrm{BF}}}}
\frac{1}{\lambda_{\textrm{BN}}}e^{-\frac{y}{\lambda_{\textrm{BN}}}}
\frac{1}{\lambda_{\textrm{NF}}}e^{-\frac{z}{\lambda_{\textrm{NF}}}}
dzdydx.
\label{Pr_NHF=0_NOMA1}
\end{align}
By performing some algebraic arrangements and invoking \cite[Eq. (3.324.1)]{TableOfIntegrals}, we can arrive at
\begin{align}
&\textrm{Pr}^{\textrm{NOMA}}\left\lbrace \textrm{NHF=0,FDF=0,NDF=1,NDN=1} \right\rbrace\nonumber\\
&=e^{-\frac{C_{\textrm{N}}^{\textrm{NOMA}}}{\lambda_{\textrm{BN}}}}\times
\left(1-e^{-\frac{ d_{\textrm{BF}}^{\alpha} \sigma_{\textrm{F}}^2 \left( 2^R-1\right) } {\lambda_{\textrm{BF}}\left( P_{\textrm{F}}-P_{\textrm{N}} \left( 2^R-1\right) \right) } } -\int_{0}^{\frac{ d_{\textrm{BF}}^{\alpha} \sigma_{\textrm{F}}^2 \left( 2^R-1\right) } {P_{\textrm{F}}-P_{\textrm{N}} \left( 2^R-1\right) }}
\frac{\psi_1\left(x \right) K_1\left(\psi_1\left(x \right) \right)  }{\lambda_{\textrm{BF}}}e^{-\frac{x}{\lambda_{\textrm{BF}}}}dx
\right) 
,
\label{Pr_NHF=0_NOMA2}
\end{align}
where $\psi_1\left(x \right) \triangleq
\sqrt{\frac{4d_{\textrm{BN}}^{\alpha} d_{\textrm{NF}}^{\alpha}  \sigma_{\textrm{F}}^2
		\left(\left( 2^R-1\right) -\frac{P_{\textrm{F}} x}{P_{\textrm{N}} x +d_{\textrm{BF}}^{\alpha}  \sigma_{\textrm{F}}^2} \right) }
	{\lambda_{\textrm{BN}} \lambda_{\textrm{NF}} \eta P_{\textrm{B}}}}
$ and $K_1(\cdot)$ denotes the first order modified Bessel function of the second kind \cite[Eq. (8.407)]{TableOfIntegrals}. In the same way, one can show that
\begin{align}
&\textrm{Pr}^{\textrm{NOMA}}\left\lbrace \textrm{FHN=0,NDF=0,FDF=1,FDN=1} \right\rbrace
\nonumber\\
&
=e^{-\frac{C_{\textrm{F}}^{\textrm{NOMA}}}{\lambda_{\textrm{BF}}}}\times\left(1-e^{-\frac{ d_{\textrm{BN}}^{\alpha} \sigma_{\textrm{N}}^2 \left( 2^R-1\right) } {\lambda_{\textrm{BN}}\left( P_{\textrm{F}}-P_{\textrm{N}} \left( 2^R-1\right) \right) } }
-\int_{0}^{\frac{ d_{\textrm{BN}}^{\alpha} \sigma_{\textrm{N}}^2 \left( 2^R-1\right) } {P_{\textrm{F}}-P_{\textrm{N}} \left( 2^R-1\right) }}
\frac{\psi_2\left(y \right) K_1\left(\psi_2\left(y \right) \right)  }{\lambda_{\textrm{BN}}}e^{-\frac{y}{\lambda_{\textrm{BN}}}}dy
 \right) 
,
\end{align}
where $\psi_2\left(y \right) \triangleq
\sqrt{\frac{4d_{\textrm{BF}}^{\alpha} d_{\textrm{NF}}^{\alpha}  \sigma_{\textrm{N}}^2
		\left(\left( 2^R-1\right) -\frac{P_{\textrm{N}} y}{P_{\textrm{F}} y +d_{\textrm{BN}}^{\alpha}  \sigma_{\textrm{N}}^2} \right) }
	{\lambda_{\textrm{BF}} \lambda_{\textrm{NF}} \eta P_{\textrm{B}}}}
$.
Next, according to \eqref{SNR_NDF}$\sim$\eqref{SNR_NDN}, \eqref{SNR_FDN}, \eqref{Beta_F} and \eqref{SNR_FHN2}, when $k \le 2^R$, $\textrm{Pr}^{\textrm{NOMA}}\left\lbrace \textrm{FHN=0,NDF=1,NDN=0,FDF=1,FDN=1}\right\rbrace=0$ holds. Otherwise, using \cite[Eq. (3.324.1)]{TableOfIntegrals} again, we can arrive at
\begin{align}
&\textrm{Pr}^{\textrm{NOMA}}\left\lbrace \textrm{FHN=0,NDF=1,NDN=0,FDF=1,FDN=1} \right\rbrace=e^{-\frac{C_{\textrm{F}}^{\textrm{NOMA}}}{\lambda_{\textrm{BF}}}}
\nonumber\\
&\times 
\left(e^{-\frac{ d_{\textrm{BN}}^{\alpha} \sigma_{\textrm{N}}^2 \left( 2^R-1\right) } {\lambda_{\textrm{BN}}\left( P_{\textrm{F}}-P_{\textrm{N}} \left( 2^R-1\right) \right) } }-e^{-\frac{d_{\textrm{BN}}^{\alpha} \sigma_{\textrm{N}}^{2}\left(2^{R}-1\right)}{\lambda_{\textrm{BN}} P_{\textrm{N}}} } 
-\int_{\frac{ d_{\textrm{BN}}^{\alpha} \sigma_{\textrm{N}}^2 \left( 2^R-1\right) } {P_{\textrm{F}}-P_{\textrm{N}} \left( 2^R-1\right) }}^{\frac{d_{\textrm{BN}}^{\alpha} \sigma_{\textrm{N}}^{2}\left(2^{R}-1\right)}{P_{\textrm{N}}}}
\frac{\psi_3\left(y \right) K_1\left(\psi_3\left(y \right) \right)  }{\lambda_{\textrm{BN}}}e^{-\frac{y}{\lambda_{\textrm{BN}}}}dy
\right),
\end{align}
where $\psi_3\left(y \right) \triangleq
\sqrt{\frac{4d_{\textrm{BF}}^{\alpha} d_{\textrm{NF}}^{\alpha}  \sigma_{\textrm{N}}^2
		\left(\left( 2^R-1\right) -\frac{P_{\textrm{N}} y}{d_{\textrm{BN}}^{\alpha}  \sigma_{\textrm{N}}^2} \right) }
	{\lambda_{\textrm{BF}} \lambda_{\textrm{NF}} \eta P_{\textrm{B}}}}
$.

\textbf{A-2: High-SNR Probability of the Events in the NOMA Framework}

As $P_\textrm{B}$ approaches to infinity, \eqref{Pr_NDF=0_NOMA} can be rewritten as
\begin{align}
&\textrm{Pr}^{\textrm{NOMA}}\left\lbrace \textrm{NDF=0} \right\rbrace=\int_{0}^{\frac{d_{\textrm{BN}}^{\alpha} \alpha_{\textrm{N}}^{2}\left(2^{R}-1\right)(1+k)}{\left( k-\left(2^{\textrm{R}}-1\right)\right)P_{\textrm{B}} }}
\frac{1}{\lambda_{\textrm{BN}}}e^{ \frac{-y}{\lambda_{\textrm{BN}}}} dy
\to
\frac{d_{\textrm{BN}}^{\alpha} \sigma_{\textrm{N}}^{2}\left(2^{R}-1\right)(1+k)}{\lambda_{\textrm{BN}}\left(k-\left(2^{\textrm{R}}-1\right)\right)} \frac{1}{P_{\textrm{B}}}.
\label{Pr_NDF=0_NOMA_HighSNR}
\end{align}
In the same way, as $P_\textrm{B}\to \infty$, one can show that
\begin{align}
\textrm{Pr}^{\textrm{NOMA}}\left\lbrace \textrm{FDF=0} \right\rbrace
\to
\frac{d_{\textrm{BF}}^{\alpha} \sigma_{\textrm{F}}^{2}\left(2^{R}-1\right)(1+k)}{\lambda_{\textrm{BF}}\left(k-\left(2^{\textrm{R}}-1\right)\right)} \frac{1}{P_{\textrm{B}}}.
\label{Pr_FDF=0_NOMA_HighSNR}
\end{align}
Similarly, for the case $k>2^R$, as $P_\textrm{B}\to \infty$, it follows from \eqref{Pr_NDF=1,NDN=0_NOMA} and \eqref{Pr_FDF=1,FDN=0_NOMA} that
\begin{align}
&\textrm{Pr}^{\textrm{NOMA}}\left\lbrace \textrm{NDF=1,NDN=0}\right\rbrace 
\to
\frac{d_{\textrm{BN}}^{\alpha} \sigma_{\textrm{N}}^{2}\left(2^{R}-1\right)(1+k)\left(k-2^{\textrm{R}}\right)}{\lambda_{\textrm{BN}}\left(k-\left(2^{\textrm{R}}-1\right)\right)} \frac{1}{P_{\textrm{B}}}.
\label{Pr_NDF=1,NDN=0_NOMA_HighSNR}
\end{align}
\begin{align}
&\textrm{Pr}^{\textrm{NOMA}}\left\lbrace \textrm{FDF=1,FDN=0}\right\rbrace 
\to
\frac{d_{\textrm{BF}}^{\alpha} \sigma_{\textrm{F}}^{2}\left(2^{R}-1\right)(1+k)\left(k-2^{\textrm{R}}\right)}{\lambda_{\textrm{BF}}\left(k-\left(2^{\textrm{R}}-1\right)\right)} \frac{1}{P_{\textrm{B}}}.
\label{Pr_FDF=1,FDN=0_NOMA_HighSNR}
\end{align}
Next, as $P_\textrm{B}\to \infty$, by using \cite[Eq. (9.6.11)]{HandbookOf} and \cite[Eq. (9.6.7)]{HandbookOf}, \eqref{Pr_NHF=0_NOMA2} can be expressed as
\begin{align}
&\textrm{Pr}^{\textrm{NOMA}}\left\lbrace \textrm{NHF=0,FDF=0,NDF=1,NDN=1} \right\rbrace
\to 
\int_{0}^{\frac{ d_{\textrm{BF}}^{\alpha} \sigma_{\textrm{F}}^2 \left( 2^R-1\right) } {P_{\textrm{F}}-P_{\textrm{N}} \left( 2^R-1\right) }}
\frac{e^{-\frac{C_{\textrm{N}}^{\textrm{NOMA}}}{\lambda_{\textrm{BN}}}-\frac{x}{\lambda_{\textrm{BF}}}}}{2\lambda_{\textrm{BF}}} \ln \left(2 \psi_{1}(x)^{-1}\right) \psi_{1}(x)^{2}
dx.
\label{Pr_NHF=0_NOMA_HighSNR1}
\end{align}
Note that $C_{\textrm{N}}^{\textrm{NOMA}}\to0$ and $\frac{ d_{\textrm{BF}}^{\alpha} \sigma_{\textrm{F}}^2 \left( 2^R-1\right) } {P_{\textrm{F}}-P_{\textrm{N}} \left( 2^R-1\right) }\to0$ hold as $P_\textrm{B}\to0$. Therefore, we can arrive at
\begin{align}
&\textrm{Pr}^{\textrm{NOMA}}\left\lbrace \textrm{NHF=0,FDF=0,NDF=1,NDN=1} \right\rbrace
\to 
\int_{0}^{\frac{ d_{\textrm{BF}}^{\alpha} \sigma_{\textrm{F}}^2 \left( 2^R-1\right) } {P_{\textrm{F}}-P_{\textrm{N}} \left( 2^R-1\right) }}
\frac{1}{\lambda_{\textrm{BF}}} \ln \left(2 \psi_{1}(x)^{-1}\right) \frac{1}{2} \psi_{1}(x)^{2}
dx.
\label{Pr_NHF=0_NOMA_HighSNR2}
\end{align}
Expanding $\psi_1(x)$ in \eqref{Pr_NHF=0_NOMA_HighSNR2} and then performing the change of variables, we can conclude that
\begin{align}
&\textrm{Pr}^{\textrm{NOMA}}\left\lbrace \textrm{NHF=0,FDF=0,NDF=1,NDN=1} \right\rbrace
\to 
\frac{d_{\textrm{BF}}^{\alpha} d_{\textrm{BN}}^{\alpha} d_{\textrm{NF}}^{\alpha} \sigma_{\textrm{F}}^{4}\left(2^{R}-1\right)(1+k)}{\lambda_{\textrm{BF}} \lambda_{\textrm{BN}} \lambda_{\textrm{NF}} \eta\left(k-\left(2^{\textrm{R}}-1\right)\right)} 
\nonumber\\
&\times\int_{0}^{1}
\ln \left(\frac{\frac{\lambda_{\textrm{BN}} \lambda_{\textrm{NF}} \eta P_{\textrm{B}}}{d_{\textrm{BN}}^{\alpha} d_{\textrm{NF}}^{\alpha}  \sigma_{\textrm{F}}^{2}}}{\left(2^{R}-1\right)-\frac{k}{1+\frac{k-\left(2^{\textrm{R}}-1\right)}{x\left(2^{R}-1\right)}}}\right)  \left( \left(2^{R}-1\right)-\frac{k}{1+\frac{k-\left(2^{\textrm{R}}-1\right)}{x\left(2^{R}-1\right)}}\right) 
dx\times\frac{1}{P_{\textrm{B}}^{2}}
.
\label{Pr_NHF=0_NOMA_HighSNR3}
\end{align}
In the same way, as $P_\textrm{B}\to\infty$, one can show that
\begin{align}
&\textrm{Pr}^{\textrm{NOMA}}\left\lbrace \textrm{FHN=0,NDF=0,FDF=1,FDN=1} \right\rbrace
\to 
\frac{d_{\textrm{BF}}^{\alpha} d_{\textrm{BN}}^{\alpha} d_{\textrm{NF}}^{\alpha} \sigma_{\textrm{N}}^{4}\left(2^{R}-1\right)(1+k)}{\lambda_{\textrm{BF}} \lambda_{\textrm{BN}} \lambda_{\textrm{NF}} \eta\left(k-\left(2^{\textrm{R}}-1\right)\right)} 
\nonumber\\
&\times\int_{0}^{1}
\ln \left(\frac{\frac{\lambda_{\textrm{BF}} \lambda_{\textrm{NF}} \eta P_{\textrm{B}}}{d_{\textrm{BF}}^{\alpha} d_{\textrm{NF}}^{\alpha}  \sigma_{\textrm{N}}^{2}}}{\left(2^{R}-1\right)-\frac{k}{1+\frac{k-\left(2^{\textrm{R}}-1\right)}{y\left(2^{R}-1\right)}}}\right)  \left( \left(2^{R}-1\right)-\frac{k}{1+\frac{k-\left(2^{\textrm{R}}-1\right)}{y\left(2^{R}-1\right)}}\right) 
dy\times\frac{1}{P_{\textrm{B}}^{2}}
,
\label{Pr_FHN=0_NOMA_HighSNR}
\end{align}
and for the case $k>2^R$, we can arrive at
\begin{align}
&\textrm{Pr}^{\textrm{NOMA}}\left\lbrace \textrm{FHN=0,NDF=1,NDN=0,FDF=1,FDN=1} \right\rbrace
\to 
\frac{d_{\textrm{BF}}^{\alpha} d_{\textrm{BN}}^{\alpha} d_{\textrm{NF}}^{\alpha} \sigma_{\textrm{N}}^{4}\left(2^{R}-1\right)^{2}(1+k)\left(k-2^{\textrm{R}}\right)^{2}}{\lambda_{\textrm{BF}} \lambda_{\textrm{BN}} \lambda_{\textrm{NF}} \eta\left(k-\left(2^{\textrm{R}}-1\right)\right)^{2}}
\nonumber\\
&\times \left\lbrace\ln \left(\sqrt{\frac{\lambda_{\textrm{BF}} \lambda_{\textrm{NF}} \eta P_{\textrm{B}}}{d_{\textrm{BF}}^{\alpha} d_{\textrm{NF}}^{\alpha} \sigma_{\textrm{N}}^{2}\left(2^{R}-1\right)}}\right) \right. 
\left.+ \frac{\left(1-2 \ln \left(\frac{k-2^{\textrm{R}}}{k-\left(2^{\textrm{R}}-1\right)}\right)\right)}{4}\right\rbrace 
\frac{1}{P_{\textrm{B}}^{2}}.
\label{Pr_FHN=0_longer_NOMA_HighSNR}
\end{align}

\textbf{A-3: DMT Analysis of the CSANC and ISANC Protocols}

\subsubsection{DMT at User F for Both the CSANC and ISANC Protocols}
Note that the expression of $P_{\textrm{out,F}}^{\textrm{CSANC}}$ is the same as that of $P_{\textrm{out,F}}^{\textrm{ISANC}}$, which means that the DMT at user F for CSANC is the same as that for ISANC. Therefore, in the following, we focus on the DMT at user F for CSANC, while the conclusions are applicable for ISANC.

As $P_{\textrm{B}}\to\infty$, we can derive from \eqref{MultiplexingGain_F} that
\begin{align}
2^{R} \to\left(\frac{P_{\textrm{B}} \lambda_{\textrm{BF}}}{d_{\textrm{BF}}^{\alpha} \sigma_{\textrm{F}}^{2}}\right)^{r_{\textrm{F}}} \to \infty.
\label{DMT_R_rF}
\end{align}
Next, we consider all the three possible setups of the scaling factor $a$ and the displacement factor $b$ as follows.
\\\emph{Case 1}: $a>1$

When $a>1$, as $P_{\textrm{B}}\to\infty$, it follows from \eqref{k} and \eqref{DMT_R_rF} that $k>2^{R}$ and 
\begin{align}
	k \rightarrow a\left(\frac{P_{\textrm{B}} \lambda_{\textrm{BF}}}{d_{\textrm{BF}}^{\alpha} \sigma_{\textrm{F}}^{2}}\right)^{r_{\textrm{F}}}.
	\label{DMT_k_Case1_rF}
\end{align}
In this case, it follows from \eqref{P_N_CSANC} and \eqref{P_F_CSANC} that $P_{\textrm{out,F}}^{\textrm{CSANC}}$ consists of two items. They are $S_1 \triangleq 
 \textrm{Pr}^{\textrm{NOMA}}\left\lbrace \textrm{FDF=0} \right\rbrace \times\left(  \textrm{Pr}^{\textrm{NOMA}}\left\lbrace \textrm{NDF=0} \right\rbrace+ \textrm{Pr}^{\textrm{NOMA}}\left\lbrace \textrm{NDF=1,NDN=0} \right\rbrace\right)$ and $S_2 \triangleq 
 \textrm{Pr}^{\textrm{NOMA}}\{\textrm{NHF=0,}$ $\textrm{FDF=0,NDF=1,NDN=1}\} $ (i.e., $P_{\textrm{out,F}}^{\textrm{CSANC}}=S_1+S_2$). To obtain the DMT at F, based on the high SNR expressions \eqref{Pr_NDF=0_NOMA_HighSNR}$\sim$\eqref{Pr_NDF=1,NDN=0_NOMA_HighSNR}, we first rewrite $S_1$ with $r_\textrm{F}$ by using \eqref{DMT_R_rF} and \eqref{DMT_k_Case1_rF} as
\begin{align}
S_1\to\frac{d_{\textrm{BF}}^{\alpha} d_{\textrm{BN}}^{\alpha} \sigma_{\textrm{F}}^{2} \sigma_{\textrm{N}}^{2} a^{2}}{\lambda_{\textrm{BF}} \lambda_{\textrm{BN}}(a-1)}\left(\frac{\lambda_{\textrm{BF}}}{d_{\textrm{BF}}^{\alpha} \sigma_{\textrm{F}}^{2}}\right)^{3 r_{\textrm{F}}} \frac{1}{P_{\textrm{B}}^{2-3 r_{\textrm{F}}}}.
\label{S_1_1}
\end{align}
Next, we consider $S_2$. Based on \eqref{Pr_NHF=0_NOMA1}, one can show that
\begin{align}
&S_{2} \leq \textrm{e}^{-\frac{C_{\textrm{N}}^{\textrm{NOMA}}}{\lambda_{\textrm{BN}}}}
\int_{0}^{\frac{d_{\textrm{BF}}^{\alpha} \sigma_{\textrm{F}}^{2}\left(2^{R}-1\right)}{P_{\textrm{F}}-P_{\textrm{N}}\left(2^{R}-1\right)}}
\frac{e^{-\frac{x}{\lambda_{\textrm{BF}}}}}{\lambda_{\textrm{BF}}} 
\left(1-
\int_{0}^{\infty}
\frac{e^{-\frac{d_{\textrm{BN}}^{\alpha} d_{\textrm{NF}}^{\alpha} \sigma_{\textrm{F}}^{2}\left(2^{R}-1\right)}{\lambda_{\textrm{NF}} \eta P_{\textrm{B}} y}-\frac{y}{\lambda_{\textrm{BN}}}}}{\lambda_{\textrm{BN}}}
dy
 \right) dx \triangleq S'_2.
\end{align}
As $P_\textrm{B}\to \infty$, using \cite[Eq. (3.324.1)]{TableOfIntegrals},\cite[Eq. (9.6.11)]{HandbookOf} and \cite[Eq. (9.6.7)]{HandbookOf}, we can arrive at
\begin{align}
S'_2&\to \frac{d_{\textrm{BF}}^{\alpha} d_{\textrm{BN}}^{\alpha} d_{\textrm{NF}}^{\alpha} \sigma_{\textrm{F}}^{4}}{\lambda_{\textrm{BF}} \lambda_{\textrm{BN}} \lambda_{\textrm{NF}} \eta}
\ln \left(\frac{\lambda_{\textrm{BN}} \lambda_{\textrm{NF}} \eta P_{\textrm{B}}}{d_{\textrm{BN}}^{\alpha} d_{\textrm{NF}}^{\alpha} \sigma_{\textrm{F}}^{2}\left(2^{R}-1\right)}\right)
\frac{\left(2^{R}-1\right)^{2}(1+k)}{\left(k-\left(2^{R}-1\right)\right) P_{\textrm{B}}^{2}}.
\label{S'_2_1}
\end{align}
Rewriting \eqref{S'_2_1} with $r_\textrm{F}$ using \eqref{DMT_R_rF} and \eqref{DMT_k_Case1_rF}, we have
\begin{align}
S'_2&\to \frac{d_{\textrm{BF}}^{\alpha} d_{\textrm{BN}}^{\alpha} d_{\textrm{NF}}^{\alpha} \sigma_{\textrm{F}}^{4} a}{\lambda_{\textrm{BF}} \lambda_{\textrm{BN}} \lambda_{\textrm{NF}} \eta(a-1)}
\ln \left(\frac{\lambda_{\textrm{BN}} \lambda_{\textrm{NF}} \eta P_{\textrm{B}}^{1-r_{\textrm{F}}}}{d_{\textrm{BN}}^{\alpha} d_{\textrm{NF}}^{\alpha} \sigma_{\textrm{F}}^{2}\left(\frac{\lambda_{\textrm{BF}}}{d_{\textrm{BF}}^{\alpha} \sigma_{\textrm{F}}^{2}}\right)^{r_{\textrm{F}}}}\right)
\left(\frac{\lambda_{\textrm{BF}}}{d_{\textrm{BF}}^{\alpha} \sigma_{\textrm{F}}^{2}}\right)^{2 r_{\textrm{F}}} \frac{1}{P_{\textrm{B}}^{2-2 r_{\textrm{F}}}}.
\label{S'_2_2}
\end{align}
Since $S_2\le S'_2$, it follows from \eqref{S_1_1} and \eqref{S'_2_2} that the DMT at F is $(2-3r_\textrm{F} )$ in Case 1.
\\\emph{Case 2}: $a=1,0<b\le1$

When $a=1,0<b\le1$, as $P_{\textrm{B}}\to\infty$, it follows from \eqref{k} and \eqref{DMT_R_rF} that $k\le2^R$ and
\begin{align}
k \rightarrow \left(\frac{P_{\textrm{B}} \lambda_{\textrm{BF}}}{d_{\textrm{BF}}^{\alpha} \sigma_{\textrm{F}}^{2}}\right)^{r_{\textrm{F}}}.
\label{DMT_k_Case2_rF}
\end{align}  
In this case, $P_{\textrm{out,F}}^{\textrm{CSANC}}=S_2+S_3$ with $S_3$ denoting $\textrm{Pr}^{\textrm{NOMA}}\left\lbrace \textrm{FDF=0} \right\rbrace \times \textrm{Pr}^{\textrm{NOMA}}\left\lbrace \textrm{NDF=0} \right\rbrace$. To obtain the DMT at F, based on the high SNR expressions \eqref{Pr_NDF=0_NOMA_HighSNR} and \eqref{Pr_FDF=0_NOMA_HighSNR}, we first rewrite $S_3$ with $r_\textrm{F}$ by using \eqref{DMT_R_rF} and \eqref{DMT_k_Case2_rF} as
\begin{equation}
S_{3} \to \frac{d_{\textrm{BF}}^{\alpha} d_{\textrm{BN}}^{\alpha} \sigma_{\textrm{F}}^{2} \sigma_{\textrm{N}}^{2}}{\lambda_{\textrm{BF}} \lambda_{\textrm{BN}} b^{2}}
\left(\frac{\lambda_{\textrm{BF}}}{d_{\textrm{BF}}^{\alpha} \sigma_{\textrm{F}}^{2}}\right)^{4 r_{\textrm{F}}} \frac{1}{P_{\textrm{B}}^{2-4 r_{\textrm{F}}}}.
\label{S_3}
\end{equation}
Note that \eqref{S'_2_1} still holds in Case 2, based on which we arrive at
\begin{align}
&S'_2\to
\frac{d_{\textrm{BF}}^{\alpha} d_{\textrm{BN}}^{\alpha} d_{\textrm{NF}}^{\alpha} \sigma_{\textrm{F}}^{4}}{\lambda_{\textrm{BF}} \lambda_{\textrm{BN}} \lambda_{\textrm{NF}} \eta b}
\ln \left( \frac{\lambda_{\textrm{BN}} \lambda_{\textrm{NF}} \eta P_{\textrm{B}}^{1-r_{\textrm{F}}}}{d_{\textrm{BN}}^{\alpha} d_{\textrm{NF}}^{\alpha} \sigma_{\textrm{F}}^{2}\left(\frac{\lambda_{\textrm{BF}}}{d_{\textrm{BF}}^{\alpha} \sigma_{\textrm{F}}^{2}}\right)^{r_{\textrm{F}}}}\right) 
\left(\frac{\lambda_{\textrm{BF}}}{d_{\textrm{BF}}^{\alpha} \sigma_{\textrm{F}}^{2}}\right)^{3 r_{\textrm{F}}} \frac{1}{P_{\textrm{B}}^{2-3 r_{\textrm{F}}}}
\label{S'_2_case2}
\end{align}
by using \eqref{DMT_R_rF} and \eqref{DMT_k_Case2_rF} again. Noting that $S_2\le S'_2$, by combining \eqref{S_3} and \eqref{S'_2_case2}, we can determine that the DMT at F is $(2-4r_\textrm{F} )$ in Case 2.
\\\emph{Case 3}: $a=1,b>1$

When $a=1,b>1$, as $P_{\textrm{B}}\to\infty$, it follows from \eqref{k} and \eqref{DMT_R_rF} that $k>2^R$ and the asymptotic expression \eqref{DMT_k_Case2_rF} with respect to $k$ still holds herein. In this case, the equation $P_{\textrm{out,F}}^{\textrm{CSANC}}=S_1+S_2$ holds as in Case 1 and the asymptotic expression of $S'_2$ (i.e., \eqref{S'_2_case2}) holds as in Case 2. Thus, we only need to rewrite $S_1$ with $r_\textrm{F}$ by using \eqref{DMT_R_rF} and \eqref{DMT_k_Case2_rF} as below:
\begin{align}
S_1\to\frac{d_{\textrm{BF}}^{\alpha} d_{\textrm{BN}}^{\alpha} \sigma_{\textrm{F}}^{2} \sigma_{\textrm{N}}^{2}}{\lambda_{\textrm{BF}} \lambda_{\textrm{BN}} b}
\left(\frac{\lambda_{\textrm{BF}}}{d_{\textrm{BF}}^{\alpha} \sigma_{\textrm{F}}^{2}}\right)^{4 r_{\textrm{F}}} \frac{1}{P_{\textrm{B}}^{2-4 r_{\textrm{F}}}}.
\label{S_1_Case3}
\end{align}
Since $S_2\le S'_2$, if follows from \eqref{S'_2_case2} and \eqref{S_1_Case3} that the DMT at F is $(2-4r_\textrm{F} )$ in Case 3.
\subsubsection{DMT at User N for the CSANC Protocol}
Considering the DMT at user N, as $P_{\textrm{B}}\to\infty$, we can derive from \eqref{MultiplexingGain_N} that
\begin{align}
2^{R} \to\left(\frac{P_{\textrm{B}} \lambda_{\textrm{BN}}}{d_{\textrm{BN}}^{\alpha} \sigma_{\textrm{N}}^{2}}\right)^{r_{\textrm{N}}} \to \infty.
\label{DMT_R_rN}
\end{align}
Next, we consider all the three possible setups of the scaling factor $a$ and the displacement factor $b$ as follows.
\\\emph{Case 1}: $a>1$

In this case, it follows from \eqref{k} and \eqref{DMT_R_rN} that $k>2^{R}$ and
\begin{align}
k \rightarrow a\left(\frac{P_{\textrm{B}} \lambda_{\textrm{BN}}}{d_{\textrm{BN}}^{\alpha} \sigma_{\textrm{N}}^{2}}\right)^{r_{\textrm{N}}}.
\label{DMT_k_Case1_rN}
\end{align}  
By using \eqref{DMT_R_rN} and \eqref{DMT_k_Case1_rN}, as $P_\textrm{B}\to \infty$, it follows from \eqref{P_N_CSANC}, \eqref{Pr_NDF=0_NOMA_HighSNR} and \eqref{Pr_NDF=1,NDN=0_NOMA_HighSNR} that
\begin{align}
&P_{\textrm{out,N}}^{\textrm{CSANC}}\to\frac{d_{\textrm{BN}}^{\alpha} \sigma_{\textrm{N}}^{2} a}{\lambda_{\textrm{BN}}}\left(\frac{\lambda_{\textrm{BN}}}{d_{\textrm{BN}}^{\alpha} \sigma_{\textrm{N}}^{2}}\right)^{2 r_{\textrm{N}}}
\frac{1}{P_{\textrm{B}}^{1-2 r_{\textrm{N}}}}.
\end{align}
Therefore, the DMT at N for CSANC is $(1-2r_\textrm{N} )$ in Case 1.
\\\emph{Case 2}: $a=1,0<b\le1$

When $a=1,0<b\le1$, it follows from \eqref{k} and \eqref{DMT_R_rN} that $k\le2^{R}$ and
\begin{align}
k \rightarrow \left(\frac{P_{\textrm{B}} \lambda_{\textrm{BN}}}{d_{\textrm{BN}}^{\alpha} \sigma_{\textrm{N}}^{2}}\right)^{r_{\textrm{N}}}.
\label{DMT_k_Case2_rN}
\end{align}  
Note that in this case $\textrm{Pr}^{\textrm{NOMA}}\left\lbrace \textrm{NDF=1,NDN=0} \right\rbrace=0$ holds. By using \eqref{DMT_R_rN} and \eqref{DMT_k_Case2_rN}, as $P_\textrm{B}\to \infty$, it follows from \eqref{P_N_CSANC} and \eqref{Pr_NDF=0_NOMA_HighSNR} that
\begin{align}
&P_{\textrm{out,N}}^{\textrm{CSANC}}\to
\frac{d_{\textrm{BN}}^{\alpha} \sigma_{\textrm{N}}^{2}}{\lambda_{\textrm{BN}} b}\left(\frac{\lambda_{\textrm{BN}}}{d_{\textrm{BN}}^{\alpha} \sigma_{\textrm{N}}^{2}}\right)^{2 r_{\textrm{N}}}
\frac{1}{P_{\textrm{B}}^{1-2 r_{\textrm{N}}}}.
\end{align}
Therefore, the DMT at N for CSANC is $(1-2r_\textrm{N} )$ in Case 2.
\\\emph{Case 3}: $a=1,b>1$

For the case $a=1,b>1$, it follows from \eqref{k} and \eqref{DMT_R_rN} that $k>2^R$ and the asymptotic expression \eqref{DMT_k_Case2_rN} with respect to $k$ still holds herein. Next, by using \eqref{DMT_R_rN} and \eqref{DMT_k_Case2_rN} again, as $P_\textrm{B}\to \infty$, it follows from \eqref{P_N_CSANC}, \eqref{Pr_NDF=0_NOMA_HighSNR} and \eqref{Pr_NDF=1,NDN=0_NOMA_HighSNR} that
\begin{align}
&P_{\textrm{out,N}}^{\textrm{CSANC}}\to
\frac{d_{\textrm{BN}}^{\alpha} \sigma_{\textrm{N}}^{2}}{\lambda_{\textrm{BN}}}\left(\frac{\lambda_{\textrm{BN}}}{d_{\textrm{BN}}^{\alpha} \sigma_{\textrm{N}}^{2}}\right)^{2 r_{\textrm{N}}}
\frac{1}{P_{\textrm{B}}^{1-2 r_{\textrm{N}}}},
\end{align}
which shows that in Case 3, the DMT at N for CSANC is still $(1-2r_\textrm{N} )$.

\subsubsection{DMT at User N for the ISANC Protocol}
The DMT at user N for the ISANC protocol can be determined by rewriting the derived high-SNR expression of $P_{\textrm{out,N}}^{\textrm{ISANC}}$ with $r_\textrm{N}$ using \eqref{DMT_R_rN}. As before, all the three possible setups of the scaling factor $a$ and the displacement factor $b$ should be considered. The detailed derivation is omitted here due to the space limit, which is similar to our analysis process to the DMT at user F for the CSANC and ISANC protocols above.

\section{}
\setcounter{equation}{0}

\textbf{B-1: Probability of the Events in the OFDMA Framework}

According to \eqref{SNR_NDF_O} and \eqref{SNR_NDN_O}, we can obtain
\begin{align}
&\textrm{Pr}^{\textrm{OFDMA}}\left\lbrace \textrm{NDN=0}\right\rbrace =1-\exp \left(-\frac{d_{\textrm{BN}}^{\alpha} \sigma_{\textrm{N}}^{2}(1-\theta)}{\lambda_{\textrm{BN}} P_{\textrm{N}}}\left(2^{\frac{R}{1-\theta}}-1\right)\right).
\label{Pr_NDN=0_OFDMA}
\end{align}
\begin{align}
\textrm{Pr}^{\textrm{OFDMA}}\left\lbrace \textrm{NDF=1,NDN=1}\right\rbrace =\exp \left( -\frac{C_{\textrm{N}}^{\textrm{OFDMA}}}{\lambda_{\textrm{BN}}}\right) .
\label{Pr_NDF=1,NDN=1_OFDMA}
\end{align}
Next, according to \eqref{SNR_FDF_O} and \eqref{SNR_FDN_O}, when $\frac{(1-\theta)}{P_{\textrm{N}}}\left(2^{\frac{R}{1-\theta}}-1\right) \leq \frac{\theta}{P_{\textrm{F}}}\left(2^{\frac{R}{\theta}}-1\right)$, we have 
$\textrm{Pr}^{\textrm{OFDMA}}\{\textrm{FDF}$ $\textrm{=1,FDN=0}\}=0$. Otherwise, one can show that
\begin{align} 
&\textrm{Pr}^{\textrm{OFDMA}}\left\lbrace \textrm{FDF=1,FDN=0}\right\rbrace=
e^{-\frac{d_{\textrm{BF}}^{\alpha} \sigma_{\textrm{F}}^{2} \theta}{\lambda_{\textrm{BF}} P_{\textrm{F}}}\left(2^{\frac{R}{\theta}}-1\right)} 
-e^{-\frac{d_{\textrm{BF}}^{\alpha} \sigma_{\textrm{F}}^{2}(1-\theta)}{\lambda_{\textrm{BF}} P_{\textrm{N}}}\left(2^{\frac{R}{1-\theta}}-1\right)}.
\label{Pr_FDF=1,FDN=0_OFDMA}
\end{align}
Furthermore, based on \eqref{SNR_NDF_O}$\sim$\eqref{Beta_N_O}, \eqref{SNR_FDF_O} and \eqref{SNR_NHF_O}, using \cite[Eq. (3.324.1)]{TableOfIntegrals} again, we have
\begin{align}
&\textrm{Pr}^{\textrm{OFDMA}}\left\lbrace \textrm{NHF=0,FDF=0,NDF=1,NDN=1}\right\rbrace \nonumber\\
&=e^{-\frac{C_{\textrm{N}}^{\textrm{OFDMA}}}{\lambda_{\textrm{BN}}}}\left(1-
e^{-\frac{d_{\textrm{BF}}^{\alpha} \sigma_{\textrm{F}}^{2} \theta}{\lambda_{\textrm{BF}} P_{\textrm{F}}}\left(2^{\frac{R}{\theta}}-1\right)}-\int_{0}^{\frac{d_{\textrm{BF}}^{\alpha} \sigma_{\textrm{F}}^{2} \theta}{P_{\textrm{F}}}\left(2^{\frac{R}{\theta}}-1\right)} \frac{e^{-\frac{x}{\lambda_{\textrm{BF}}}}}{\lambda_{\textrm{BF}}}  \psi_{4}(x) K_{1}\left(\psi_{4}(x)\right)dx\right)
,
\label{Pr_NHF=0,FDF=0,NDF=1,NDN=1_OFDMA2}
\end{align}
where $\psi_{4}(x)=\sqrt{\frac{4\left(\left(2^{\frac{R}{\theta}}-1\right)-\frac{P_{\textrm{F}} x}{d_{\textrm{BF}}^{\alpha} \sigma_{\textrm{F}}^{2} \theta}\right)}
	{\frac{\lambda_{\textrm{BN}} \lambda_{\textrm{NF}} \eta P_{\textrm{B}}}{d_{\textrm{BN}}^{\alpha} d_{\textrm{NF}}^{\alpha} \sigma_{\textrm{F}}^{2} \theta}}}$. As stated in Section III, user N and user F are peers in ISAOC such that the analysis of the two users is absolutely symmetrical. As thus, by replacing $d_{\textrm{BN}}$, $d_{\textrm{BF}}$, $
\lambda_{\textrm{BN}}$, $\lambda_{\textrm{BF}}$, $\sigma_{\textrm{N}}$, $\sigma_{\textrm{F}}$, $P_{\textrm{N}}$, $P_{\textrm{F}}$, $C_{\textrm{N}}^{\textrm{OFDMA}}$, $C_{\textrm{F}}^{\textrm{OFDMA}}$ and $\theta$ with $d_{\textrm{BF}}$, $d_{\textrm{BN}}$, $\lambda_{\textrm{BF}}$, $
\lambda_{\textrm{BN}}$, $\sigma_{\textrm{F}}$, $\sigma_{\textrm{N}}$, $P_{\textrm{F}}$, $P_{\textrm{N}}$, $C_{\textrm{F}}^{\textrm{OFDMA}}$, $C_{\textrm{N}}^{\textrm{OFDMA}}$ and $\left(1-\theta \right)$, respectively, the analytical expressions of $\textrm{Pr}^{\textrm{OFDMA}}\left\lbrace \textrm{FDF=0}\right\rbrace$, $\textrm{Pr}^{\textrm{OFDMA}}\left\lbrace \textrm{FDF=1,FDN=1}\right\rbrace$ and $\textrm{Pr}^{\textrm{OFDMA}}\left\lbrace \textrm{FHN=0,NDN=0,FDF=1,FDN=1}\right\rbrace$ can be directly obtained from equations \eqref{Pr_NDN=0_OFDMA}, \eqref{Pr_NDF=1,NDN=1_OFDMA} and \eqref{Pr_NHF=0,FDF=0,NDF=1,NDN=1_OFDMA2}, respectively. For example, from \eqref{Pr_NDF=1,NDN=1_OFDMA}, we can determine that
\begin{align}
\textrm{Pr}^{\textrm{OFDMA}}\left\lbrace \textrm{FDF=1,FDN=1}\right\rbrace =\exp \left( -\frac{C_{\textrm{F}}^{\textrm{OFDMA}}}{\lambda_{\textrm{BF}}}\right) .
\label{Pr_FDF=1,FDN=1_OFDMA}
\end{align}

\textbf{B-2: High-SNR Probability of the Events in the OFDMA Framework}

We define $\rho\in(0,1)$ as the percentage of the power resources that is allocated to user F (i.e., $\rho\triangleq P_\textrm{F}/P_\textrm{B}$). As $P_\textrm{B}$ approaches to infinity, \eqref{Pr_NDN=0_OFDMA} leads to
\begin{align}
&\textrm{Pr}^{\textrm{OFDMA}}\left\lbrace \textrm{NDN=0}\right\rbrace 
\to \frac{d_{\textrm{BN}}^{\alpha} \sigma_{\textrm{N}}^{2}(1-\theta)}{\lambda_{\textrm{BN}}(1-\rho)}\left(2^{\frac{R}{1-\theta}}-1\right) \frac{1}{P_{\textrm{B}}}.
\label{Pr_NDN=0_OFDMA_HighSNR}
\end{align}
Similarly, as $P_\textrm{B}\to\infty$, it follows from \eqref{Pr_NDF=1,NDN=1_OFDMA} and \eqref{C_N_OFDMA} that
\begin{align}
&\left( 1-\textrm{Pr}^{\textrm{OFDMA}}\left\lbrace \textrm{NDF=1,NDN=1}\right\rbrace\right)  \to
\frac{C_{\textrm{N}}^{\textrm{OFDMA}}}{\lambda_{\textrm{BN}}} \nonumber\\
&\to
\begin{cases}
&\frac{d_{\textrm{BN}}^{\alpha} \sigma_{\textrm{N}}^{2}(1-\theta)}{\lambda_{\textrm{BN}}(1-\rho)}\left(2^{\frac{R}{1-\theta}}-1\right) \frac{1}{P_{\textrm{B}}},
\textrm{if } \frac{ \left(1-\theta \right)  } {P_{\textrm{N}}} \left( 2^{\frac{R}{\left(1-\theta \right)}}-1\right) > \frac{\theta } {P_{\textrm{F}}} \left( 2^{\frac{R}{\theta} }-1\right),\\
&\frac{ d_{\textrm{BN}}^{\alpha} \sigma_{\textrm{N}}^2 \theta } {\lambda_{\textrm{BN}}\rho} \left( 2^{\frac{R}{\theta} }-1\right)\frac{1}{P_{\textrm{B}}},
\textrm{if }\frac{ \left(1-\theta \right)  } {P_{\textrm{N}}} \left( 2^{\frac{R}{\left(1-\theta \right)}}-1\right) \le \frac{\theta } {P_{\textrm{F}}} \left( 2^{\frac{R}{\theta} }-1\right). 
\end{cases}
\label{Pr_NDF=1,NDN=1_OFDMA_HighSNR}
\end{align}
In the same way, for the case $\frac{(1-\theta)}{P_{\textrm{N}}}\left(2^{\frac{R}{1-\theta}}-1\right) > \frac{\theta}{P_{\textrm{F}}}\left(2^{\frac{R}{\theta}}-1\right)$, as $P_\textrm{B}\to\infty$, we can derive from \eqref{Pr_FDF=1,FDN=0_OFDMA} that
\begin{align} 
&\textrm{Pr}^{\textrm{OFDMA}}\left\lbrace \textrm{FDF=1,FDN=0}\right\rbrace
\to \frac{d_{\textrm{BF}}^{\alpha} \sigma_{\textrm{F}}^{2}}{\lambda_{\textrm{BF}}}\left(\frac{(1-\theta)}{(1-\rho)}\left(2^{\frac{R}{1-\theta}}-1\right)-\frac{\theta}{\rho}\left(2^{\frac{R}{\theta}}-1\right)\right) \frac{1}{P_{\textrm{B}}}.
\end{align}
Next, following a similar derivation process as that in our derivation of equation \eqref{Pr_NHF=0_NOMA_HighSNR3}, as $P_\textrm{B}\to\infty$, we can derive from \eqref{Pr_NHF=0,FDF=0,NDF=1,NDN=1_OFDMA2} that
\begin{align}
&\textrm{Pr}^{\textrm{OFDMA}}\left\lbrace \textrm{NHF=0,FDF=0,NDF=1,NDN=1}\right\rbrace \nonumber\\
&\to\frac{d_{\textrm{BF}}^{\alpha} d_{\textrm{BN}}^{\alpha} d_{\textrm{NF}}^{\alpha} \sigma_{\textrm{F}}^{4} \theta^{2}}{\lambda_{\textrm{BF}} \lambda_{\textrm{BN}} \lambda_{\textrm{NF}} \eta \rho}\left(2^{\frac{R}{\theta}}-1\right)^{2}
\left\{\ln \left(\sqrt{\frac{\lambda_{\textrm{BN}} \lambda_{\textrm{NF}} \eta P_{\textrm{B}}}{d_{\textrm{BN}}^{\alpha} d_{\textrm{NF}}^{\alpha} \sigma_{\textrm{F}}^{2} \theta\left(2^{\frac{R}{\theta}}-1\right)}}\right)+\frac{1}{4}\right\} \frac{1}{P_{\textrm{B}}^{2}}
 .
\label{Pr_NHF=0,FDF=0,NDF=1,NDN=1_OFDMA_HighSNR}
\end{align}
As stated Appendix B-1, by replacing $d_{\textrm{BN}}$, $d_{\textrm{BF}}$, $
\lambda_{\textrm{BN}}$, $\lambda_{\textrm{BF}}$, $\sigma_{\textrm{N}}$, $\sigma_{\textrm{F}}$, $P_{\textrm{N}}$, $P_{\textrm{F}}$, $C_{\textrm{N}}^{\textrm{OFDMA}}$, $C_{\textrm{F}}^{\textrm{OFDMA}}$, $\rho$ and $\theta$ with $d_{\textrm{BF}}$, $d_{\textrm{BN}}$, $\lambda_{\textrm{BF}}$, $
\lambda_{\textrm{BN}}$, $\sigma_{\textrm{F}}$, $\sigma_{\textrm{N}}$, $P_{\textrm{F}}$, $P_{\textrm{N}}$, $C_{\textrm{F}}^{\textrm{OFDMA}}$, $C_{\textrm{N}}^{\textrm{OFDMA}}$, $\left(1-\rho \right)$ and $\left(1-\theta \right)$, respectively, the asymptotic expressions of $\textrm{Pr}^{\textrm{OFDMA}}\left\lbrace \textrm{FDF=0}\right\rbrace$, $(1-\textrm{Pr}^{\textrm{OFDMA}}\left\lbrace \textrm{FDF=1,FDN=1}\right\rbrace)$ and $\textrm{Pr}^{\textrm{OFDMA}}\left\lbrace \textrm{FHN=0,NDN=0,FDF=1,FDN=1}\right\rbrace$ can be directly obtained from \eqref{Pr_NDN=0_OFDMA_HighSNR}, \eqref{Pr_NDF=1,NDN=1_OFDMA_HighSNR} and \eqref{Pr_NHF=0,FDF=0,NDF=1,NDN=1_OFDMA_HighSNR}, respectively. For example, as $P_\textrm{B}\to\infty$, from \eqref{Pr_NDN=0_OFDMA_HighSNR}, we can determine that
\begin{align}
\textrm{Pr}^{\textrm{OFDMA}}\left\lbrace \textrm{FDF=0}\right\rbrace \to\frac{d_{\textrm{BF}}^{\alpha} \sigma_{\textrm{F}}^{2} \theta}{\lambda_{\textrm{BF}} \rho }\left(2^{\frac{R}{\theta}}-1\right)\frac{1}{P_{\textrm{B}}}.
\label{Pr_FDF=0_OFDMA_HighSNR}
\end{align}

\textbf{B-3: Proof of Proposition 7}

Since the analysis of the two users is absolutely symmetrical for ISAOC, we can directly obtain the DMT at user N according to the derived DMT at user F. Therefore, we focus on the DMT at user F in the following. According to \eqref{P_F_ISAOC}, $P_{\textrm{out,F}}^{\textrm{ISAOC}}$ consists of two items. We denote them as $S_7\triangleq\textrm{Pr}^{\textrm{OFDMA}}\left\lbrace \textrm{FDF=0} \right\rbrace \times \left(1- \textrm{Pr}^{\textrm{OFDMA}}\left\lbrace \textrm{NDF=1,NDN=1} \right\rbrace\right)$ and $S_8\triangleq \textrm{Pr}^{\textrm{OFDMA}}
\{\textrm{NHF=0,FDF=0,NDF=}$ $\textrm{1,NDN=1}\}$ (i.e., $P_{\textrm{out,F}}^{\textrm{ISAOC}}=S_7+S_8$). Next, to determine the DMT at user F, we rewrite the two items with $r_{\textrm{F}}$ as in Appendix A-3 as follows: According to \eqref{Pr_NDF=1,NDN=1_OFDMA_HighSNR}, \eqref{Pr_FDF=0_OFDMA_HighSNR} and \eqref{DMT_R_rF}, as $P_\textrm{B}\to\infty$, for the case $\frac{ \left(1-\theta \right)  } {P_{\textrm{N}}} \left( 2^{\frac{R}{\left(1-\theta \right)}}-1\right) > \frac{\theta } {P_{\textrm{F}}} \left( 2^{\frac{R}{\theta} }-1\right)$, we have
\begin{align}
S_7\to
\frac{d_{\textrm{BF}}^{\alpha} d_{\textrm{BN}}^{\alpha} \sigma_{\textrm{F}}^{2} \sigma_{\textrm{N}}^{2}\theta(1-\theta) }{\lambda_{\textrm{BF}} \lambda_{\textrm{BN}} \rho(1-\rho)}
\left(\frac{\lambda_{\textrm{BF}}}{d_{\textrm{BF}}^{\alpha} \sigma_{\textrm{F}}^{2}}\right)^{\frac{r_{\textrm{F}}}{\theta\left(1-\theta \right) }}
\frac{1}{P_{\textrm{B}}^{2-\frac{r_{\textrm{F}}}{\theta\left(1-\theta \right) }}}. 
\label{S_7_case1}
\end{align}
On the other hand, for the case $\frac{ \left(1-\theta \right)  } {P_{\textrm{N}}} \left( 2^{\frac{R}{\left(1-\theta \right)}}-1\right) \le \frac{\theta } {P_{\textrm{F}}} \left( 2^{\frac{R}{\theta} }-1\right)$, one can show that
\begin{align}
S_7
\to\frac{d_{\textrm{BF}}^{\alpha} d_{\textrm{BN}}^{\alpha} \sigma_{\textrm{F}}^{2} \sigma_{\textrm{N}}^{2} \theta^{2}}{\lambda_{\textrm{BF}} \lambda_{\textrm{BN}} \rho^{2} }
\left(\frac{\lambda_{\textrm{BF}}}{d_{\textrm{BF}}^{\alpha} \sigma_{\textrm{F}}^{2}}\right)^{\frac{2 r_{\textrm{F}}}{\theta}}
\frac{1}{P_{\textrm{B}}^{2-\frac{2 r_{\textrm{F}}}{\theta}}}. 
\label{S_7_case2}
\end{align}
Next, according to \eqref{Pr_NHF=0,FDF=0,NDF=1,NDN=1_OFDMA_HighSNR} and \eqref{DMT_R_rF}, as $P_\textrm{B}\to\infty$, we can rewrite $S_8$ with $r_{\textrm{F}}$ as
\begin{align}
&S_8\to\frac{d_{\textrm{BF}}^{\alpha} d_{\textrm{BN}}^{\alpha} d_{\textrm{NF}}^{\alpha} \sigma_{\textrm{F}}^{4} \theta^{2}}{\lambda_{\textrm{BF}} \lambda_{\textrm{BN}} \lambda_{\textrm{NF}} \eta \rho}
\left(\frac{\lambda_{\textrm{BF}}}{d_{\textrm{BF}}^{\alpha} \sigma_{\textrm{F}}^{2}}\right)^{\frac{2 r_{\textrm{F}}}{\theta}}
\left\{\ln \left(\sqrt{\frac{\lambda_{\textrm{BN}} \lambda_{\textrm{NF}} \eta P_{\textrm{B}}^{1-\frac{ r_{\textrm{F}}}{\theta}}}{d_{\textrm{BN}}^{\alpha} d_{\textrm{NF}}^{\alpha} \sigma_{\textrm{F}}^{2} \theta\left(\frac{\lambda_{\textrm{BF}}}{d_{\textrm{BF}}^{\alpha} \sigma_{\textrm{F}}^{2}}\right)^{\frac{ r_{\textrm{F}}}{\theta}}}}\right)+\frac{1}{4}\right\} \frac{1}{P_{\textrm{B}}^{2-\frac{2 r_{\textrm{F}}}{\theta}}}
.
\label{S_8}
\end{align}
Combining \eqref{S_7_case1} \eqref{S_7_case2} and \eqref{S_8}, we can determine that the DMT at user F can be given by
$\min\left\lbrace2-\frac{r_{\textrm{F}}}{\theta\left( 1-\theta\right) }, 2-\frac{2r_{\textrm{F}}}{\theta }\right\rbrace$ when
$\frac{ \left(1-\theta \right)  } {P_{\textrm{N}}} \left( 2^{\frac{R}{\left(1-\theta \right)}}-1\right) > \frac{\theta } {P_{\textrm{F}}} \left( 2^{\frac{R}{\theta} }-1\right)$. Otherwise, the DMT at user F can be expressed as $\left( 2-\frac{2r_{\textrm{F}}}{\theta} \right)$, which completes the proof.

\ifCLASSOPTIONcaptionsoff
\newpage
\fi
\bibliographystyle{IEEEtran}
\bibliography{TWCR}

\begin{thebibliography}{10}
\providecommand{\url}[1]{#1}
\csname url@samestyle\endcsname
\providecommand{\newblock}{\relax}
\providecommand{\bibinfo}[2]{#2}
\providecommand{\BIBentrySTDinterwordspacing}{\spaceskip=0pt\relax}
\providecommand{\BIBentryALTinterwordstretchfactor}{4}
\providecommand{\BIBentryALTinterwordspacing}{\spaceskip=\fontdimen2\font plus
\BIBentryALTinterwordstretchfactor\fontdimen3\font minus
  \fontdimen4\font\relax}
\providecommand{\BIBforeignlanguage}[2]{{%
\expandafter\ifx\csname l@#1\endcsname\relax
\typeout{** WARNING: IEEEtran.bst: No hyphenation pattern has been}%
\typeout{** loaded for the language `#1'. Using the pattern for}%
\typeout{** the default language instead.}%
\else
\language=\csname l@#1\endcsname
\fi
#2}}
\providecommand{\BIBdecl}{\relax}
\BIBdecl

\bibitem{Dai15CommMag}
L.~{Dai}, B.~{Wang}, Y.~{Yuan}, S.~{Han}, C.~{I}, and Z.~{Wang},
  ``Non-orthogonal multiple access for {5G}: solutions, challenges,
  opportunities, and future research trends,'' \emph{IEEE Commun. Mag.},
  vol.~53, no.~9, pp. 74--81, Sep. 2015.

\bibitem{Islam17CommSur}
S.~M.~R. {Islam}, N.~{Avazov}, O.~A. {Dobre}, and K.~{Kwak}, ``Power-domain
  non-orthogonal multiple access ({NOMA}) in {5G} systems: Potentials and
  challenges,'' \emph{IEEE Commun. Surveys Tuts.}, vol.~19, no.~2, pp.
  721--742, Secondquarter 2017.

\bibitem{Saito13VTC}
Y.~{Saito}, Y.~{Kishiyama}, A.~{Benjebbour}, T.~{Nakamura}, A.~{Li}, and
  K.~{Higuchi}, ``Non-orthogonal multiple access ({NOMA}) for cellular future
  radio access,'' in \emph{IEEE VTC Spring}, Jun. 2013, pp. 1--5.

\bibitem{Benjebbour13ISPACS}
A.~{Benjebbour}, Y.~{Saito}, Y.~{Kishiyama}, A.~{Li}, A.~{Harada}, and
  T.~{Nakamura}, ``Concept and practical considerations of non-orthogonal
  multiple access ({NOMA}) for future radio access,'' in \emph{IEEE ISPACS},
  Nov. 2013, pp. 770--774.

\bibitem{Sadia18ELEKTRO}
H.~{Sadia}, M.~{Zeeshan}, and S.~A. {Sheikh}, ``Performance analysis of
  downlink power domain {NOMA} under fading channels,'' in \emph{ELEKTRO}, May
  2018, pp. 1--6.

\bibitem{DingZhiguo15CommuLet}
Z.~{Ding}, M.~{Peng}, and H.~V. {Poor}, ``Cooperative non-orthogonal multiple
  access in {5G} systems,'' \emph{IEEE Commun. Lett.}, vol.~19, no.~8, pp.
  1462--1465, Aug. 2015.

\bibitem{huang_ComMagz17}
J.~Huang, C.~Xing, and C.~Wang, ``Simultaneous wireless information and power
  transfer: Technologies, applications, and research challenges,'' \emph{IEEE
  Commun. Mag.}, vol.~55, no.~11, pp. 26--32, Nov. 2017.

\bibitem{Perera_ComSurTu18}
T.~D. {Ponnimbaduge Perera}, D.~N.~K. {Jayakody}, S.~K. {Sharma},
  S.~{Chatzinotas}, and J.~{Li}, ``Simultaneous wireless information and power
  transfer ({SWIPT}): Recent advances and future challenges,'' \emph{IEEE
  Commun. Surveys Tuts.}, vol.~20, no.~1, pp. 264--302, Firstquarter 2018.

\bibitem{Hu__ComSurTu18}
J.~{Hu}, K.~{Yang}, G.~{Wen}, and L.~{Hanzo}, ``Integrated data and energy
  communication network: A comprehensive survey,'' \emph{IEEE Commun. Surveys
  Tuts.}, vol.~20, no.~4, pp. 3169--3219, Fourthquarter 2018.

\bibitem{Krikidis_ComMagz14}
I.~Krikidis, S.~Timotheou, S.~Nikolaou, G.~Zheng, D.~W.~K. Ng, and R.~Schober,
  ``Simultaneous wireless information and power transfer in modern
  communication systems,'' \emph{IEEE Commun. Mag.}, vol.~52, no.~11, pp.
  104--110, Nov. 2014.

\bibitem{Varshney_ISIT’08}
L.~R. Varshney, ``Transporting information and energy simultaneously,'' in
  \emph{Proc. IEEE ISIT}, Toronto, Canada, Jul. 2008, pp. 1612--1616.

\bibitem{Grover_ISIT’10}
P.~Grover and A.~Sahai, ``Shannon meets {T}esla: Wireless information and power
  transfer,'' in \emph{Proc. IEEE ISIT}, Austin, TX, Jun. 2010, pp. 2363--2367.

\bibitem{Nasir_TWC’13}
A.~A. {Nasir}, X.~{Zhou}, S.~{Durrani}, and R.~A. {Kennedy}, ``Relaying
  protocols for wireless energy harvesting and information processing,''
  \emph{IEEE Trans. Wireless Commun.}, vol.~12, no.~7, pp. 3622--3636, Jul.
  2013.

\bibitem{liu2013TWC}
L.~{Liu}, R.~{Zhang}, and K.~{Chua}, ``Wireless information transfer with
  opportunistic energy harvesting,'' \emph{IEEE Trans. Wireless Commun.},
  vol.~12, no.~1, pp. 288--300, Jan. 2013.

\bibitem{Krikidis14TC}
I.~{Krikidis}, ``Simultaneous information and energy transfer in large-scale
  networks with/without relaying,'' \emph{IEEE Trans. Commun.}, vol.~62, no.~3,
  pp. 900--912, Mar. 2014.

\bibitem{Liu16JSAComm}
Y.~{Liu}, Z.~{Ding}, M.~{Elkashlan}, and H.~V. {Poor}, ``Cooperative
  non-orthogonal multiple access with simultaneous wireless information and
  power transfer,'' \emph{IEEE J. Select. Areas Commun.}, vol.~34, no.~4, pp.
  938--953, Apr. 2016.

\bibitem{YinghuiYe17ICC}
Y.~{Ye}, Y.~{Li}, D.~{Wang}, and G.~{Lu}, ``Power splitting protocol design for
  the cooperative {NOMA} with {SWIPT},'' in \emph{Proc. of IEEE ICC}, 2017, pp.
  1--5.

\bibitem{Do18SigTelCom}
T.~N. {Do} and B.~{An}, ``Optimal sum-throughput analysis for downlink
  cooperative {SWIPT} {NOMA} systems,'' in \emph{Proc. of SigTelCom}, Jan.
  2018, pp. 85--90.

\bibitem{Do17ICC}
N.~T. {Do}, D.~{Benevides da Costa}, T.~Q. {Duong}, and B.~{An}, ``Transmit
  antenna selection schemes for {MISO-NOMA} cooperative downlink transmissions
  with hybrid {SWIPT} protocol,'' in \emph{IEEE ICC}, May. 2017, pp. 1--6.

\bibitem{Hedayati18TVT}
M.~{Hedayati} and I.~{Kim}, ``On the performance of {NOMA} in the two-user
  {SWIPT} system,'' \emph{IEEE Trans. Veh. Technol.}, vol.~67, no.~11, pp.
  11\,258--11\,263, Nov. 2018.

\bibitem{Zheng03TIT}
L.~{Zheng} and D.~N.~C. {Tse}, ``Diversity and multiplexing: a fundamental
  tradeoff in multiple-antenna channels,'' \emph{IEEE Trans. Inf. Theory},
  vol.~49, no.~5, pp. 1073--1096, May 2003.

\bibitem{Laneman04TIT}
J.~N. {Laneman}, D.~N.~C. {Tse}, and G.~W. {Wornell}, ``Cooperative diversity
  in wireless networks: Efficient protocols and outage behavior,'' \emph{IEEE
  Trans. Inf. Theory}, vol.~50, no.~12, pp. 3062--3080, Dec. 2004.

\bibitem{Ding12TWC}
H.~{Ding}, J.~{Ge}, D.~B. {da Costa}, and Z.~{Jiang}, ``Two birds with one
  stone: Exploiting direct links for multiuser two-way relaying systems,''
  \emph{IEEE Trans. Wireless Commun.}, vol.~11, no.~1, pp. 54--59, Jan. 2012.

\bibitem{TableOfIntegrals}
I.~S. Gradshteyn and I.~M. Ryzhik, \emph{Table of Integrals, Series, and
  Products, 7th ed.}\hskip 1em plus 0.5em minus 0.4em\relax San Diego, CA:
  Academic, 2007.

\bibitem{HandbookOf}
M.~Abramowitz and I.~A. Stegun, \emph{Handbook of Mathematical Functions with
  Formulas, Graphs, and Mathematical Tables}, New York, 1972.

\end{thebibliography}
\end{document}